\documentclass[british,english,aps,manuscript]{revtex4-1}
\usepackage[T1]{fontenc}
\usepackage[latin9]{inputenc}
\usepackage{color}
\usepackage{amsmath}
\usepackage{amssymb}

\makeatletter
 
 \@ifundefined{textcolor}{}
 {%
   \definecolor{BLACK}{gray}{0}
   \definecolor{WHITE}{gray}{1}
   \definecolor{RED}{rgb}{1,0,0}
   \definecolor{GREEN}{rgb}{0,1,0}
   \definecolor{BLUE}{rgb}{0,0,1}
   \definecolor{CYAN}{cmyk}{1,0,0,0}
   \definecolor{MAGENTA}{cmyk}{0,1,0,0}
   \definecolor{YELLOW}{cmyk}{0,0,1,0}
 }

\makeatother

\usepackage{babel}
\begin{document}
\selectlanguage{british}%
\global\long\global\long\global\long\def\bra#1{\mbox{\ensuremath{\langle#1|}}}
\global\long\global\long\global\long\def\ket#1{\mbox{\ensuremath{|#1\rangle}}}
\global\long\global\long\global\long\def\bk#1#2{\mbox{\ensuremath{\ensuremath{\langle#1|#2\rangle}}}}
\global\long\global\long\global\long\def\kb#1#2{\mbox{\ensuremath{\ensuremath{\ensuremath{|#1\rangle\!\langle#2|}}}}}

\selectlanguage{english}%

\title{A universal framework for entanglement detection}

\author{Micha\l{} Oszmaniec}

\email{oszmaniec@cft.edu.pl}

\author{and Marek Ku\'{s}}

\address{Center for Theoretical Physics, Polish Academy of Sciences, Al. Lotników
32/46, 02-668 Warszawa}
\begin{abstract}
We construct nonlinear multiparty entanglement measures for distinguishable
particles, bosons and fermions. In each case properties of an entanglement
measures are related to the decomposition of the suitably chosen representation
of the relevant symmetry group onto irreducible components. In the
case of distinguishable particles considered entanglement measure
reduces to the well-known many particle concurrence. We prove that
our entanglement criterion is sufficient and necessary for pure states
living in both finite and infinite dimensional spaces. We generalize
our entanglement measures to mixed states by the convex roof extension
and give a non trivial lower bound of thus obtained generalized concurrence. 
\end{abstract}

\pacs{03.67.Mn, 03.65.Fd}

\maketitle

\section{Introduction\label{sec:Introduction}}

Entanglement is one of the features of quantum systems that makes
them different from their classical counterparts. Even since the invention
of this concept \cite{EPR,Schrod} there has been an ongoing debate
how to define precisely and quantify entanglement for various physical
systems. Entanglement is usually identified with correlations in the
composite quantum system that are stronger then any correlations that
can be exhibited by any classical system. In our article we elaborate
the method of entanglement detection based on identification of entangled
states with particular orbits of underlying group of local transformation
\cite{classical quantum states,Kotowscy,On detection}. For the another
approach to entanglement which makes use of the formalism of commuting
subalgebras of the algebra of all quantum observables see \cite{bennatti1,bennatti2}.

The general idea is as follows. We consider a Hilbert space $\mathcal{H}$
of a quantum system and a group $K$ of local, i.e. correlation-preserving
transformations, represented on $\mathcal{H}$ irreducibly as a subgroup
of the full unitary group $U\left(\mathcal{H}\right)$. Precise forms
of $\mathcal{H}$, $K$ and the representation vary and depend upon
a given physical situation. Because the global phase factor of the
wave function is irrelevant for physical applications, one considers
the corresponding action of $K$ on the complex projective space $\mathbb{P}\mathcal{H}$
rather then on $\mathcal{H}$ itself. Having introduced this language
entanglement can be defined as the property of orbits of the group
$K$ acting in $\mathbb{P}\mathcal{H}$ and any measure of entanglement
should be an invariant of $K$. We present a construction and a general
computational scheme for one particular invariant of the action of
$K$ on $\mathbb{P}\mathcal{H}$ which can be used for detection of
entanglement as it is non negative and vanishes exactly on the the
set of states that are coherent or, in other words, {}``the most
classical'' \cite{Uncertainty rel,Perelomov,Klyacko ent}. In the
current paper we focus on three concrete cases: entanglement of finite
number of distinguishable, bosonic or fermionic particles. We analyze
entanglement in situations when Hilbert spaces considered can be infinite
dimensional \cite{eisert infinite dimensions}.

What is, obviously, more interesting and at the same time more demanding
is to quantify entanglement of mixed states for all three cases, or
at least, discriminate between separable and entangled states. In
principle, having a good measure of entanglement for pure states we
can extend it to mixed ones by the so called convex roof extension
utilizing the fact that mixed separable states are convex combinations
of pure ones. However, since the convex roof construction of an entanglement
measure requires optimization over all convex pure-state decompositions
of a given mixed density matrix, the procedure is not effective, or
at least computationally demanding. What might be helpful in discriminating
separable and mixed density matrices are various estimates of so constructed
convex roof measures. We will show a unified procedure for constructing
useful lower bounds for the obtained measures in all three cases. 

In Section \ref{sec:Nonlinear-entanglement-witness} we briefly present
the general construction of nonlinear entanglement measure in finite
dimensional setting with the usage of representation-theoretic tools.
In Section \ref{sec:Explicit-expressions-for} we apply the developed
methods of discriminating entangled and non entangled pure states
in all three cases of distinguishable particles, bosons and fermions.
Number of particles as well as the dimensions of a single particle
Hilbert spaces are arbitrary but finite. In the case of distinguishable
particles our entanglement measure reduces to the well known multiparty
concurrence \cite{Wooters,Mintert1} and we decided to keep this name
calling it a generalized concurrence. Section \ref{sec:Generalization-to-infinite}
is devoted to generalization of results from Section \ref{sec:Explicit-expressions-for}
to systems with arbitrary finite number of particles but with infinite-dimensional
single-particle Hilbert spaces. We prove that entanglement criteria
that we present in Section \ref{sec:Explicit-expressions-for} still
hold in the infinite-dimensional setting. In Section \ref{sec:Entanglement-detection-for}
we extend our entanglement measures to the general mixed states via
the convex roof extension. We provide a systematic way to find lower
bounds for generalized concurrences for fermionic and bosonic systems
starting from any lower bounds for the concurrence for multiparty
distinguishable particles. These lower bounds give a sufficient conditions
of entanglement for a mixed fermionic or bosonic state. Relevant mathematical
details that we omit during the main discussion are given in the Appendices.

\section{Nonlinear entanglement measures \label{sec:Nonlinear-entanglement-witness}}

Let us briefly remind the construction of nonlinear entanglement measure
for multiparty pure states space presented in \cite{Kotowscy}. It
is based on an obvious observation that entanglement, or more general,
quantum correlations do not change under local transformations. Such
local transformations form a (Lie) group $K$ acting in the space
of states of the system hence states with the same correlation properties
({}``equally entangled'') belong to the same orbit of the group
$K$, i.e. to the set of states which can be obtained form a particular
one by applying all operations of the group $K$. In particular, non-correlated
(non-entangled) states form a single particular orbit of $K$.

\subsection{Entanglement and orbits of local groups}

The notion of a {}``local transformation'' and, consequently, the
structure of the group $K$ depends on the situation at hand. In the
case of $L$ distinguishable particles, the Hilbert space of states
is the tensor product of Hilbert spaces of single particles $\mathcal{H}_{1},\ldots,\mathcal{H}_{L}$
of dimensions $\ensuremath{N_{1},\ldots,N_{L}}$ which we conveniently
identify with the complex spaces of the same dimensions, 
\begin{equation}
\mathcal{H}_{d}=\mathcal{H}_{1}\otimes\cdots\otimes\mathcal{H}_{L}=\mathbb{C}^{N_{1}}\otimes\ldots\otimes\mathbb{C}^{N_{L}}={\color{red}\bigotimes_{i=1}^{L}\mathbb{C}^{N_{i}}}
\end{equation}
In this case the group $K$ of local transformations leaving correlations
invariant is the direct product of special unitary group $SU(N_{k})$
each acting independently in the respective one-particle space $\mathcal{H}_{k}=\mathbb{C}^{N_{k}}$,
\begin{gather}
K=\mathrm{SU}(N_{1})\times\mathrm{SU}(N_{2})\times\ldots\times\mathrm{SU}(N_{L})={\color{red}\times_{i=1}^{L}SU(N_{i})},\nonumber \\
\Pi^{d}(U_{1}\ldots,U_{L})\ket{\psi_{1}}\otimes\cdots\otimes\ket{\psi_{L}}=U_{1}\ket{\psi_{1}}\otimes\cdots\otimes U_{L}\ket{\psi_{L}}\label{eq:dist}
\end{gather}
In the notation above we made explicit that the group $K$ acts on
$\mathcal{H}$ \textit{via} its particular representation $\Pi^{d}$,
defined here by its action on simple tensors. For simplicity we assume
that the particles are identical (albeit, as stated above, distinguishable),
hence their Hilbert space of states are the same, $N_{1}=\cdots=N_{S}$,
hence
\[
\mathcal{H}_{d}=\bigotimes^{L}\left(\mathbb{C}^{N}\right),\, K=\times^{L}\big(\mathrm{SU}(N)\big)\,.
\]
 For indistinguishable particles the situation differs. The appropriate
Hilbert space of the whole system is no longer the full tensor product
of one-particle Hilbert spaces but rather its symmetric (for bosons)
or antisymmetric (for fermions) part 
\begin{gather}
\mathcal{H}_{b}=\mathrm{Sym}^{L}\left(\mathbb{C}^{N}\right)=\mathbb{C}^{N}\vee\cdots\vee\mathbb{C}^{N},\\
\mathcal{H}_{f}=\bigwedge^{L}\left(\mathbb{C}^{N}\right)=\mathbb{C}^{N}\wedge\cdots\wedge\mathbb{C}^{N},
\end{gather}
where $\vee$ and $\wedge$ denote, respectively, the symmetric and
antisymmetric part of the full tensor product. To keep symmetry intact,
the local group $K$ must consist of {}``diagonal'' actions of the
$\mathrm{SU}(N)$ group, i.e. actions of the same unitary operator
in each one-particle space, 
\begin{gather}
K=\mathrm{SU}(N),\quad\Pi^{b}(U)(\ket{\psi_{1}}\vee\cdots\vee\ket{\psi_{S}})=U\ket{\psi_{1}}\vee\cdots\vee U\ket{\psi_{S}},\label{bos}\\
K=\mathrm{SU}(N),\quad\Pi^{f}(U)(\ket{\psi_{1}}\wedge\cdots\wedge\ket{\psi_{S}})=U\ket{\psi_{1}}\wedge\cdots\wedge U\ket{\psi_{S}},\label{eq:ferm}
\end{gather}
where we denoted the appropriate representations of $K$ by $\Pi^{b}$
and $\Pi^{f}$.

From the mathematical point of view all three cases of a) distinguishable
particles (Eq. (\ref{eq:dist})), b) bosons (Eq. (\ref{bos})), and
c) fermions (Eq (\ref{eq:ferm})) are instances of the same scheme:
a compact (semi)simple group $K$ acts via irreducible representation
$\Pi$ on a Hilbert space $\mathcal{H}$ (i.e. there are no proper
subspaces of $\mathcal{H}$ preserved by $\Pi$). Each irreducible
representation of a compact semisimple group is uniquely determined
by the so called highest weight vector in the representation space
$\mathcal{H}$ (see Appendix A). In what follows, unless othervise
specified, all representation we consided will be irreduicible. From
the physical point of view it is more appropriate to consider action
of $K$ on the projective space $\mathbb{P}\mathcal{H}$ rather than
on $\mathcal{H}$ itself, as in the physical interpretation of vectors
from $\mathcal{H}$ their phase does not play a role and we use only
vectors normalized to unity. The space $\mathbb{P}\mathcal{H}$ is
the space of different complex directions in $\mathcal{H}$, each
determined by a normalized vector $\ket{\psi}$ . The group $K$ acts
naturally on $\mathbb{P}\mathcal{H}$ :

\[
\tilde{\Pi}(k)\left(\left[\ket{\psi}\right]\right)=\left[\Pi(k)\ket{\psi}\right],
\]
where $\mathcal{H}\ni\ket{\psi}\mapsto[\psi]\in\mathbb{P}\mathcal{H}$
is the mapping that associates to the unit vector $\ket{\psi}$ the
complex direction passing through it. In the above introduced language
the set of most classical (coherent or {}``non-entangled'') states
is the orbit of $K$ through the highest weight state $\left[\ket{\psi_{0}}\right]$
\cite{classical quantum states,Kotowscy,Perelomov}. Mathematically,
this orbit can be identified as the orbit of Perelomov's generalized
coherent states for the representation $\Pi$ of the group $K$, that
are ``closest to classical'' \cite{Perelomov}, 

\begin{equation}
\mathcal{O}_{0}=\left\{ \left[\ket{\psi}\right]=\left[\Pi(k)\ket{\psi_{0}}\right]|\, k\in K\right\} .\label{eq:coherent}
\end{equation}
This definition can be motivated in a threefold way. Firstly, for
the case of distinguishable particles (see below), one recovers the
standard separable states. Secondly, states that belong to this class
minimalise the $K$- invariant uncertainty of the state $\ket{\psi}$:

\[
\mathrm{Var}\left(\ket{\psi}\right)=\sum_{i}\left(\bk{\psi}{X_{i}^{2}|\psi}-\bk{\psi}{X_{i}|\psi}^{2}\right)\,,
\]
where the sum is over generators of the Lie algebra of the group $K$
(see \cite{Klyacko ent} for application of $\mathrm{Var}\left(\ket{\psi}\right)$
in entanglement theory and \cite{Puri} for application of {}``minimal
uncertainty coherent states'' in quantum optics). The third important
feature of considered classes of states is that they are appear naturally
when studding classical limits of certain models steaming from quantum
optics \cite{Puri} or condensed matter physics \cite{Chaos}.

For the case of distinguishable particles (\ref{eq:dist}), generalized
coherent states are precisely separable states,

\begin{equation}
\mathcal{O}_{sep}=\left\{ \left[\ket{\psi_{1}}\otimes\ket{\psi_{2}}\otimes\ldots\otimes\ket{\psi_{L}}\right]|\,\ket{\psi_{i}}\in\mathcal{H}_{i}\right\} \,\,.
\end{equation}
One checks that:

\begin{gather}
\mathcal{O}_{b}=\left\{ \left[\ket{\phi}\otimes\ket{\phi}\otimes\ldots\otimes\ket{\phi}\right]|\,\ket{\phi}\in\mathbb{C}^{N}\right\} \,\text{and}\\
\mathcal{O}_{f}=\left\{ \left[\ket{\phi_{1}}\wedge\ket{\phi_{2}}\wedge\ldots\wedge\ket{\phi_{L}}\right]|\,\ket{\phi_{i}}\in\mathbb{C}^{N},\,\bk{\phi_{i}}{\phi_{j}}=\delta_{ij}\right\} \,,
\end{gather}
are {}``coherent states'' for system of respectively $L$ bosons
(\ref{bos}) and $L$ fermions (\ref{eq:ferm}). The notion of entanglement
for bosons or fermions is not well defined as corresponding Hilbert
spaces $\mathcal{H}_{b}$ and $\mathcal{H}_{f}$ lack the tensor product
structure. Nevertheless, we prefer to call states belonging to $\mathcal{O}_{b}$
and $\mathcal{O}_{f}$ as {}``least entangled'' bosonic and fermionic
states. Note that $\mathcal{O}_{b}$ and $\mathcal{O}_{f}$  consist
of simplest tensors available in $\mathcal{H}_{b}$ and $\mathcal{H}_{f}$
respectively. What is more, these are exactly the sets of pure {}``separable''
states for bosons and fermions analyzed in \cite{Eckert}. Classes
of states $\mathcal{O}_{b}$ and $\mathcal{O}_{f}$ are also interesting
from the practical point of view. For $N=2$ states from $\mathcal{O}_{b}$
are exactly celebrated spin coherent states \cite{Puri}. On the other
hand, $\mathcal{O}_{f}$ is, for general $N$, a very important class
of variational states in condensed matter physics.

\subsection{Characterization of orbits of non-entangled states and generalized
concurrence }

It is now clear that identification of a state $\left[\ket{\psi}\right]$
as a non-entangled one is equivalent to checking whether it belongs
to the orbit of the local group $K$ through the highest weight vector.
A constructive way of checking this fact was given in the paper by
Lichtenstein \cite{lichteinstein}. To present it let us go back for
a moment to our general setting (for the relevant definitions consult
Appendix A). Let

\[
K\ni k\rightarrow\Pi^{\lambda_{0}}(k)\in U\left(\mathcal{H}^{\lambda_{0}}\right),
\]
be a unitary representation of $K$ characterized by the highest weight
$\lambda_{0}$ (we wrote $\mathcal{H}^{\lambda_{0}}$ instead of $\mathcal{H}$
to indicate the parameter encoding the representation). Let us introduce
auxiliary unitary representation of $K$ on the symmetric tensor product
$\mathcal{H}^{\lambda_{0}}\vee\mathcal{H}^{\lambda_{0}}$ 

\begin{equation}
K\ni k\rightarrow\Pi^{\lambda_{0}}(k)\otimes\Pi^{\lambda_{0}}(k)\in U\left(\mathrm{Sym}^{2}\left(\mathcal{H}^{\lambda_{0}}\right)\right).\label{eq:power}
\end{equation}
In general $\mathcal{H}^{\lambda_{0}}\vee\mathcal{H}^{\lambda_{0}}$
decomposes onto irreducible representations of $K$ :

\begin{equation}
\mathrm{Sym}^{2}\left(\mathcal{H}^{\lambda_{0}}\right)\approx\mathcal{H}^{2\lambda_{0}}\oplus\bigoplus_{\beta\neq2\lambda_{0}}\mathcal{H}^{\beta},
\end{equation}
where $\mathcal{H}^{2\lambda_{0}}$ is the representation of the highest
weight $2\lambda_{0}$ (one can show that there is only one representation
of this kind in the above sum \cite{Klyacko ent}) and the sum on
the right side is over other irreducible representations that appear
in the decomposition of $\mathrm{Sym}^{2}\left(\mathcal{H}^{\lambda_{0}}\right)$.
The announced result of Lichtenstein \cite{lichteinstein} states
that $\left[\ket{\psi}\right]$ is a coherent state if in and only
if $\ket{\psi}\ket{\psi}\in\mathcal{H}^{2\lambda_{0}}$. We can write
this result in the equivalent form

\begin{equation}
\left[\ket{\psi}\right]\in\mathcal{O}_{0}\Longleftrightarrow\bra{\psi}\otimes\bra{\psi}\mathbb{I}\otimes\mathbb{I}-\mathbb{P}^{2\lambda_{0}}\ket{\psi}\otimes\ket{\psi}=0\,,\label{eq:class criterion}
\end{equation}
where $\mathbb{P}^{2\lambda_{0}}$ is the orthogonal projector onto
$\mathcal{H}^{2\lambda_{0}}$ and $\mathbb{I}$ stands for the identity
operator on $\mathcal{H}^{\lambda_{0}}$. Theorem of Lichtenstein
written in this form can be used to construct the nonlinear entanglement
measure, the generalized concurrence which we define by the following
expression 
\begin{equation}
C\left(\ket{\psi}\right)=\sqrt{\bra{\psi}\otimes\bra{\psi}\mathbb{I}\otimes\mathbb{I}-\mathbb{P}^{2\lambda_{0}}\ket{\psi}\otimes\ket{\psi}}\,.\label{eq:concurrence}
\end{equation}
One easily checks that $C\left(\ket{\psi}\right)$ is non negative
and vanishes exactly for coherent states. Moreover, it is also $K$
invariant. These two conditions allow us to treat $C\left(\ket{\psi}\right)$
as an indicator of entanglement. 

Although the above construction works only for compact group represented
in the finite dimensional Hilbert space in Section \ref{sec:Generalization-to-infinite}
we will generalize the concurrence also to systems of distinguishable
particles, fermions or bosons described in infinite dimensional Hilbert
spaces.

\section{Explicit expressions for generalized concurrences for pure states\label{sec:Explicit-expressions-for}}

It possible to compute a detailed form of the projector operator $\mathbb{P}^{2\lambda_{0}}$
acting $\mathrm{Sym}^{2}\left(\mathcal{H}^{\lambda_{0}}\right)$ for
the case of distinguishable particles as well as for bosons and fermions.
The proofs for formulas for $\mathbb{P}^{2\lambda_{0}}$ rely on representation
theory and are given in Appendix (\ref{sec:Apppendix 1}). We obtain
explicit form of the function $C\left(\ket{\psi}\right)$. For the
case of distinguishable particles we recover the well-known multiparty
concurrence \cite{Wooters,Mintert1}. Results we get for bosons and
fermions are generalization of the previous paper of one of the authors
\cite{Kotowscy}. The main advantage of our generalization lies in
the fact that it reveals a strong connection between concurrences
for non-distinguishable particles with the one defined for distinguishable
particles. This connection allows for the {}``physical interpretation''
of $C\left(\ket{\psi}\right)$ for non-distinguishable particles in
terms of the reduced density matrices of the state $\ket{\psi}.$
In Section \ref{sec:Entanglement-detection-for} we use this connection
to the problem of detection of mixed entangled states. We obtain non-trivial
lower bounds for concurrences for bosons and fermions from any lower
bound for the multiparty concurrence. In the Section \ref{sec:Generalization-to-infinite}
we prove that the formulas for concurrences obtained in this part
hold also in the infinite dimensional setting.

\subsection{Distinguishable particles\label{sub:Distinguishable-particles}}

In the case of $L$ distinguishable particles we have $\mathcal{H}^{\lambda_{0}}=\mathcal{H}_{d}=\bigotimes_{i=1}^{i=L}\mathcal{H}_{i}$
and the symmetry group $K=\times_{i=1}^{i=L}SU(N)$ acts on it (\ref{eq:dist}).
It is easy to extract give a compact form of $\mathbb{P}^{2\lambda_{0}}$.
Let us first introduce some notation: 

\begin{equation}
\mathcal{H}_{d}\otimes\mathcal{H}_{d}=\left(\bigotimes_{i=1}^{i=L}\mathcal{H}_{i}\right)\otimes\left(\bigotimes_{i=1'}^{i=L'}\mathcal{H}_{i}\right),\label{eq:notation dist}
\end{equation}
where $L=L'$ and we decided to label spaces from the second copy
of the total space with primes in order to avoid ambiguity. Action
of $K$ on $\mathrm{Sym}^{2}\left(\mathcal{H}_{d}\right)$ is given
by the restriction to the symmetric (with respect to the interchange
of copies of $\mathcal{H}_{d}$) tensors of the action defined on
$\mathcal{H}_{d}\otimes\mathcal{H}_{d}$ (\ref{eq:power}). Let us
also introduce the symmetrization operators $\mathbb{P}_{ii'}^{+}:\mathcal{H}_{d}\otimes\mathcal{H}_{d}\rightarrow\mathcal{H}_{d}\otimes\mathcal{H}_{d}$
that project onto the subspace of $\mathcal{H}_{d}\otimes\mathcal{H}_{d}$
completely symmetric under interchange spaces $i$ and $i'$ (one
can define anti-symmetrization operators $\mathbb{P}_{ii'}^{+}$ in
the analogous way). Under introduced notation we have the closed expression
for the projector operator $\mathbb{P}^{2\lambda_{0}}$

\begin{equation}
\mathbb{P}^{2\lambda_{0}}=\mathbb{P}_{d}=\mathbb{P}_{11'}^{+}\circ\mathbb{P}_{22'}^{+}\circ\ldots\circ\mathbb{P}_{LL'}^{+}\,.\label{eq:entanglement wittness dist}
\end{equation}

We can now write down explicitly our entanglement measure for distinguishable
particles. For $\ket{\psi}\in\mathcal{H}_{d}$ we have

\begin{equation}
C_{d}\left(\ket{\psi}\right)=\sqrt{\bra{\psi}\bra{\psi}\mathbb{I}\otimes\mathbb{I}-\mathbb{P}_{11'}^{+}\circ\mathbb{P}_{22'}^{+}\circ\ldots\circ\mathbb{P}_{LL'}^{+}\ket{\psi}\ket{\psi}}\,,\label{eq:conc disting}
\end{equation}
where subscript $d$ stands from distinguishable particles. Expression
above is, up to a multiplicative factor, the well-known multipartite
concurrence \cite{Wooters,Mintert1}.

\subsection{Bosons \label{sub:Bosons}}

Hilbert space describing $L$ bosonic particles has the structure
$\mathcal{H}^{\lambda_{0}}=\mathcal{H}_{b}=\mathrm{Sym}^{L}\left(\mathcal{H}\right)$,
where $\mathcal{H}\approx\mathbb{C}^{N}$. The symmetry group represented
in this space is $K=SU(N)$ (\ref{bos}). We embed $\mathrm{Sym}^{L}\left(\mathcal{H}\right)$
in the Hilbert space of $L$ identical distinguishable particles,

\begin{equation}
\mathrm{Sym}^{L}\left(\mathcal{H}\right)\subset\mathcal{H}_{1}\otimes\ldots\otimes\mathcal{H}_{L},
\end{equation}
where $\mathcal{H}_{i}\approx\mathcal{H}$. We have the analogous
embedding of $\mathrm{Sym}^{L}\left(\mathcal{H}\right)\vee\mathrm{Sym}^{L}\left(\mathcal{H}\right)$,

\begin{equation}
\mathrm{Sym}^{L}\left(\mathcal{H}\right)\vee\mathrm{Sym}^{L}\left(\mathcal{H}\right)\subset\left(\bigotimes_{i=1}^{i=L}\mathcal{H}_{i}\right)\otimes\left(\bigotimes_{i=1'}^{i=L'}\mathcal{H}_{i}\right)=\mathcal{H}_{d}\otimes\mathcal{H}_{d},\label{eq:notation bos}
\end{equation}
where, as before, $L=L'$. Let $\mathbb{P}_{\left\{ 1,\ldots,L\right\} }^{\mathrm{sym}}:\mathcal{H}_{d}\otimes\mathcal{H}_{d}\rightarrow\mathcal{H}_{d}\otimes\mathcal{H}_{d}$
be the projector onto the subspace of $\mathcal{H}_{d}\otimes\mathcal{H}_{d}$
which is completely symmetric with respect to interchange of spaces
labeled by indices from the set $\left\{ 1,2,\ldots,L\right\} $.
We define $\mathbb{P}_{\left\{ 1',\ldots,L'\right\} }^{\mathrm{sym}}$
in the analogous way. Under the above notation operator $\mathbb{P}^{2\lambda_{0}}$
takes the form:

\begin{equation}
\mathbb{P}^{2\lambda_{0}}=\mathbb{P}_{b}=\left(\mathbb{P}_{11'}^{+}\circ\mathbb{P}_{22'}^{+}\circ\ldots\circ\mathbb{P}_{LL'}^{+}\right)\left(\mathbb{P}_{\left\{ 1,\ldots,L\right\} }^{\mathrm{sym}}\circ\mathbb{P}_{\left\{ 1',\ldots,L'\right\} }^{\mathrm{sym}}\right),\label{eq:entanglement wittness bos}
\end{equation}
where it is understood that $\mathbb{P}^{2\lambda_{0}}$ acts on the
space $\mathcal{H}_{d}\otimes\mathcal{H}_{d}$ (see (\ref{eq:notation bos})).
Operators $\mathbb{P}_{ii'}^{+}$ are the same as in the previous
section. Let us note that we may write 

\begin{equation}
\left.\mathbb{P}_{b}\right|_{\mathrm{Sym}^{L}\left(\mathcal{H}\right)\otimes\mathrm{Sym}^{L}\left(\mathcal{H}\right)}=\mathbb{P}_{11'}^{+}\circ\mathbb{P}_{22'}^{+}\circ\ldots\circ\mathbb{P}_{LL'}^{+},\label{eq:bosons final}
\end{equation}
as for any $\ket{\Psi}\in\mathrm{Sym}^{L}\left(\mathcal{H}\right)\otimes\mathrm{Sym}^{L}\left(\mathcal{H}\right)$
we have $\left(\mathbb{P}_{\left\{ 1,\ldots,L\right\} }^{\mathrm{sym}}\otimes\mathbb{P}_{\left\{ 1',\ldots,L'\right\} }^{\mathrm{sym}}\right)\ket{\Psi}=\ket{\Psi}$.
Entanglement measure for bosonic particles takes the same form as
for distinguishable particles. For $\ket{\psi}\in\mathrm{Sym}^{L}\left(\mathcal{H}\right)$
we have

\begin{equation}
C_{b}\left(\ket{\psi}\right)=\sqrt{\bra{\psi}\bra{\psi}\mathbb{I}\otimes\mathbb{I}-\mathbb{P}_{11'}^{+}\circ\mathbb{P}_{22'}^{+}\circ\ldots\circ\mathbb{P}_{LL'}^{+}\ket{\psi}\ket{\psi}}\,,\label{eq:conc bos}
\end{equation}
where subscript $b$ stands for bosons.

\subsection{Fermions\label{sub:Fermions}}

Hilbert space describing $L$ fermionic particles is $\mathcal{H}^{\lambda_{0}}=\mathcal{H}_{f}=\bigwedge^{L}\left(\mathcal{H}\right)$,
where $\mathcal{H}\approx\mathbb{C}^{N}$. The symmetry group is again
$K=SU(N)$ (see (\ref{eq:ferm})). Just like in the case of bosons
(see (\ref{eq:notation bos})) we have 
\begin{gather}
\bigwedge^{L}\left(\mathcal{H}\right)\subset\mathcal{H}_{1}\otimes\ldots\otimes\mathcal{H}_{L},\label{eq:not ferm}\\
\,\bigwedge^{L}\left(\mathcal{H}\right)\vee\bigwedge^{L}\left(\mathcal{H}\right)\subset\left(\bigotimes_{i=1}^{i=L}\mathcal{H}_{i}\right)\otimes\left(\bigotimes_{i=1'}^{i=L'}\mathcal{H}_{i}\right)=\mathcal{H}_{d}\otimes\mathcal{H}_{d}\,.
\end{gather}
By $\mathbb{P}_{\left\{ 1,\ldots,L\right\} }^{a\mathrm{sym}}:\mathcal{H}_{d}\otimes\mathcal{H}_{d}\rightarrow\mathcal{H}_{d}\otimes\mathcal{H}_{d}$
we denote the projector onto the subspace of $\mathcal{H}_{d}\otimes\mathcal{H}_{d}$
which is completely asymmetric with respect to interchange of spaces
labeled by indices from the set $\left\{ 1,2,\ldots,L\right\} $.
We define $\mathbb{P}_{\left\{ 1',\ldots,L'\right\} }^{\mathrm{asym}}$
in the analogous way. Under this notation we get

\begin{equation}
\mathbb{P}^{2\lambda_{0}}=\mathbb{P}_{f}=\mbox{\ensuremath{\alpha}}\left(\mathbb{P}_{11'}^{+}\circ\mathbb{P}_{22'}^{+}\circ\ldots\circ\mathbb{P}_{LL'}^{+}\right)\left(\mathbb{P}_{\left\{ 1,\ldots,L\right\} }^{\mathrm{asym}}\circ\mathbb{P}_{\left\{ 1',\ldots,L'\right\} }^{\mathrm{asym}}\right),\label{eq:entanglement wittness bos-1}
\end{equation}
where $\alpha=\mbox{\ensuremath{\frac{2^{L}}{L+1}}}$ and it is understood
that $\mathbb{P}^{2\lambda_{0}}$ acts on the space $\mathcal{H}_{d}\otimes\mathcal{H}_{d}$.
 In analogy to the case of bosons we have 

\[
\left.\mathbb{P}_{f}\right|_{\mathrm{\bigwedge}^{L}\left(\mathcal{H}\right)\otimes\mathrm{\bigwedge}^{L}\left(\mathcal{H}\right)}=\mbox{\ensuremath{\alpha}}\mathbb{P}_{11'}^{+}\circ\mathbb{P}_{22'}^{+}\circ\ldots\circ\mathbb{P}_{LL'}^{+},
\]
since for any $\ket{\Psi}\in\mathrm{\bigwedge}^{L}\left(\mathcal{H}\right)\otimes\mathrm{\bigwedge}^{L}\left(\mathcal{H}\right)$
we have $\left(\mathbb{P}_{\left\{ 1,\ldots,L\right\} }^{a\mathrm{sym}}\circ\mathbb{P}_{\left\{ 1',\ldots,L'\right\} }^{a\mathrm{sym}}\right)\ket{\Psi}=\ket{\Psi}$.
Generalized concurrence for a fermionic state $\ket{\psi}\in\mathrm{\bigwedge}^{L}\left(\mathcal{H}\right)$
reads

\begin{equation}
C_{f}\left(\ket{\psi}\right)=\sqrt{\bra{\psi}\bra{\psi}\mathbb{I}\otimes\mathbb{I}-\alpha\mathbb{P}_{11'}^{+}\circ\mathbb{P}_{22'}^{+}\circ\ldots\circ\mathbb{P}_{LL'}^{+}\ket{\psi}\ket{\psi}}\,,\label{eq:conc ferm}
\end{equation}
where subscript $f$ stands for fermions.

\subsection{Physical interpretation of generalized concurrences}

Expressions for $C_{d}$, $C_{b}$ and $C_{f}$ depend only upon $\bra{\psi}\bra{\psi}\mathbb{P}_{11'}^{+}\circ\mathbb{P}_{22'}^{+}\circ\ldots\circ\mathbb{P}_{LL'}^{+}\ket{\psi}\ket{\psi}$.
One can show \cite{Agu=00015Bciak} that for arbitrary $L$-particle
states the following expression holds,

\begin{equation}
\bra{\psi}\bra{\psi}\mathbb{P}_{11'}^{+}\circ\mathbb{P}_{22'}^{+}\circ\ldots\circ\mathbb{P}_{LL'}^{+}\ket{\psi}\ket{\psi}=2^{-L}\left(\sum_{k}\mathrm{tr}\left(\rho_{k}^{2}\right)+2\right)\,,\label{eq:phys form}
\end{equation}
where the summation is over all different $2^{L}-2$ proper subsystems
of $L$-partcile systems and $\rho_{k}$ is the reduced density matrix
describing the particular subsystem. Notice that the expression (\ref{eq:phys form})
is also valid for bosons and fermions because we can formally embed
bosonic and fermionic Hilbert spaces in $\bigotimes^{L}\left(\mathbb{C}^{N}\right)$. 

Although in our reasoning we care only whether a given multiparty
pure state is {}``classical '' or not, it is tempting to ask what
are the {}``maximally entangled'' states corresponding to measures
$C_{d}$, $C_{b}$ and $C_{f}$ in each of three considered contexts.
Equation (\ref{eq:phys form}) enables us to to formally answer to
this question. Clearly, $C_{d}\left(\ket{\psi}\right)$, $C_{b}\left(\ket{\psi}\right)$
and $C_{f}\left(\ket{\psi}\right)$ will be maximal once for each
proper subsystem $k$ the corresponding reduced density matrix will
be maximally mixed. For the case of distinguishable particles states
$\ket{\psi}$ satisfying this condition are called {}``absolutely
maximally entangled states''. The problem of deciding whether for
a given $L$ and $N$ such states at all exist is in general unsolved.
Therefore, one cannot hope for an easy characterization of states
that maximize $C_{d}$, $C_{b}$ or $C_{f}$ (or equivalently, minimize
(\ref{eq:phys form}) once $\ket{\psi}\in\mathcal{H}_{d},\,\mathcal{H}_{b}$
or $\mathcal{H}_{f}$ respectively). Nevertheless, the characterization
of {}``absolutely maximally entangled'' bosonic and fermionic states
is certainly an interesting open problem.

\section{Generalization to infinite dimensional systems\label{sec:Generalization-to-infinite}}

In this section we extend the concept of concurrence for the infinite
dimensional setting. We first make a few technical remarks about infinite
dimensional setting. In the rest of the section we prove that we can
generalize the concept of concurrence introduced in previous two sections
also to infinite dimensional Hilbert spaces describing arbitrary finite
number of distinguishable particles, fermions or bosons. We prove
that criteria for entanglement given by expressions (\ref{eq:conc disting}),
(\ref{eq:conc bos}) and (\ref{eq:conc ferm}) are also valid in the
infinite dimensional setting.

Separable Hilbert space $\mathcal{H}$ is, by definition, a Hilbert
space in which it is possible to chose a countable basis. Almost all
Hilbert spaces that occur in physics are separable \cite{reed-simon}.
Examples include all finite dimensional Hilbert spaces or the space
space of square integrable (with respect to the Lebesgue measure)
functions on $\mathbb{R}^{d}$, $L^{2}\left(\mathbb{R}^{d},dx\right)$.
In this section we consider, unless we indicate otherwise, only general
separable Hilbert spaces. The space of pure states of a quantum system
is a projective space $\mathbb{P}\mathcal{H}$ which we identify with
the collection of rank one orthogonal projectors acting on $\mathcal{H}$.
The projective space $\mathbb{P}\mathcal{H}$ is metric space with
respect to the Hilbert\textendash{}Schmidt metric \cite{reed-simon}.
That is, for $\left[\ket{\psi}\right],\left[\ket{\phi}\right]\in\mathbb{P}\mathcal{H}$
we have

\begin{equation}
\mathrm{d}\left(\left[\ket{\psi}\right],\left[\ket{\phi}\right]\right)=\sqrt{\mathrm{tr}\left(\left(\kb{\psi}{\psi}-\kb{\phi}{\phi}\right)^{2}\right)}=\sqrt{2\left(1-\left|\bk{\psi}{\phi}\right|^{2}\right)}\,,\label{eq:metric structure}
\end{equation}
where $\mathrm{d}\left(\cdot,\,\cdot\right)$ denotes the metric.
The projective space endowed with the above metric is a complete metric
space, i.e. every Cauchy sequence of elements from $\mathbb{P}\mathcal{H}$
converge.

\subsection{Distinguishable particles\label{sub:Distinguishable-particles-inf}}

We first study entanglement of $L$ distinguishable particles, described
by the Hilbert space $\mathcal{H}_{d}=\bigotimes_{i=1}^{i=L}\mathcal{H}_{i}$
, where single particle Hilbert spaces $\mathcal{H}_{i}$ are in general
infinite dimensional. The notion of the tensor product of infinite
dimensional Hilbert spaces involves, by definition, taking into account
tensors having infinite rank, i.e. tensors that cannot we written
as a finite combination of elements of the form $\ket{\psi_{1}}\otimes\ket{\psi_{2}}\otimes\ldots\otimes\ket{\psi_{L}}$.
This phenomenon does not occur when dimensions of single particle
Hilbert spaces are finite. The set of separable states consists of
states having the form of simple tensors from $\mathcal{H}_{d}$,

\begin{equation}
\mathcal{O}_{\mathrm{sep}}=\left\{ \left[\ket{\psi_{1}}\otimes\ket{\psi_{2}}\otimes\ldots\otimes\ket{\psi_{L}}\right]|\,\ket{\psi_{i}}\in\mathcal{H}_{i}\right\} \,.
\end{equation}
One can identify $\mathcal{O}_{\mathrm{sep}}$ with the orbit of $K=\mathrm{U}(\mathcal{H}_{1})\times\mathrm{U}(\mathcal{H}_{2})\times\ldots\times\mathrm{U}(\mathcal{H}_{L})$
through one exemplary separable state $\left[\ket{\psi}_{\mathrm{sep}}\right]$.
The main difference with the finite dimensional setting is that the
group $K$ is not a Lie group not to mention it is compact or semisimple.
Therefore, methods of representation theory of Lie group cannot be
applied to get result of the form (\ref{eq:class criterion}). Nevertheless
we argue that the following holds,

\begin{equation}
\left[\ket{\psi}\right]\in\mathcal{O}_{\mathrm{sep}}\Longleftrightarrow\bra{\psi}\otimes\bra{\psi}\mathbb{I}\otimes\mathbb{I}-\mathbb{P}_{11'}^{+}\circ\mathbb{P}_{22'}^{+}\circ\ldots\circ\mathbb{P}_{LL'}^{+}\ket{\psi}\otimes\ket{\psi}=0,\label{eq:dist inf dim}
\end{equation}
where $\mathbb{P}_{ii'}^{+}:\mathcal{H}_{d}\otimes\mathcal{H}_{d}\rightarrow\mathcal{H}_{d}\otimes\mathcal{H}_{d}$
are the symmetrization operators defined as in Part \ref{sub:Distinguishable-particles}.
Note that (\ref{eq:dist inf dim}) implies that nonzero $C_{d}\left(\ket{\psi}\right)$
defined as in (\ref{eq:conc disting}) detects entangled pure states.
In order to prove (\ref{eq:dist inf dim}) we first observe that $\bra{\psi}\bra{\psi}\mathbb{I}\otimes\mathbb{I}-\mathbb{P}_{11'}^{+}\circ\mathbb{P}_{22'}^{+}\circ\ldots\circ\mathbb{P}_{LL'}^{+}\ket{\psi}\ket{\psi}=0$
for separable states. Therefore, we only need to prove the inverse
implication. Let us denote by $\mathcal{O}_{sep}^{i}$ the set of
states that are separable with respect to the bipartition $\mathcal{H}_{d}=\mathcal{H}_{i}\otimes\left(\bigotimes_{j\neq i}\mathcal{H}_{j}\right)$
. That is,

\begin{equation}
\mathcal{O}_{sep}^{i}=\left\{ \left[\ket{\psi}\otimes\ket{\phi}\right]|\,\ket{\psi}\in\mathcal{H}_{i}\,,\,\ket{\phi}\in\left(\bigotimes_{j\neq i}\mathcal{H}_{j}\right)\right\} .\label{eq:sep part}
\end{equation}
 One checks that $\left[\ket{\psi}\right]\in\mathcal{O}_{sep}$ if
and only if $\left[\ket{\psi}\right]\in\mathcal{O}_{sep}^{i}$ for
all $i=1,\ldots,L$ (in other words $\left[\ket{\psi}\right]$ is
separable with respect to any bipartition $\mathcal{H}_{d}=\mathcal{H}_{i}\otimes\left(\bigotimes_{j\neq i}\mathcal{H}_{j}\right)$).
Note that in order not to complicate the notation we the abuse the
notation of the tensor product in (\ref{eq:sep part}) (we do not
respect the order of terms in the tensor product). For the proof of
above statement see the Appendix B. We can now prove that $\bra{\psi}\bra{\psi}\mathbb{I}\otimes\mathbb{I}-\mathbb{P}_{11'}^{+}\otimes\mathbb{P}_{22'}^{+}\otimes\ldots\otimes\mathbb{P}_{LL'}^{+}\ket{\psi}\ket{\psi}=0$
implies that $\left[\ket{\psi}\right]$ is separable. Assume that
$\left[\ket{\psi}\right]$ is entangled. By the discussion above $\left[\ket{\psi}\right]$
is non-separable with respect to some bipartition $\mathcal{H}_{i_{0}}\otimes\left(\bigotimes_{j\neq i_{0}}\mathcal{H}_{j}\right)$.
We write the Schmidt decomposition of $\ket{\psi}$ with respect to
this bipartition \cite{eisert infinite dimensions},

\begin{equation}
\ket{\psi}=\sum_{l}\lambda_{l}\ket{\psi_{l}}\otimes\ket{\phi_{l}},\label{eq:shmidt form}
\end{equation}
where $\ket{\psi_{l}}\in\mathcal{H}_{i_{0}}$, $\ket{\phi_{l}}\in\left(\bigotimes_{j\neq i_{0}}\mathcal{H}_{j}\right)$
and $\bk{\psi_{i}}{\psi_{j}}=\bk{\phi_{i}}{\phi_{j}}=\delta_{ij}$.
Moreover, we fix the normalization of the sate by setting $\sum_{i}\left|\lambda_{i}\right|^{2}=1$.
We have:

\[
\bra{\psi}\bra{\psi}\mathbb{I}\otimes\mathbb{I}-\mathbb{P}_{11'}^{+}\circ\mathbb{P}_{22'}^{+}\circ\ldots\circ\mathbb{P}_{LL'}^{+}\ket{\psi}\ket{\psi}\geq\bra{\psi}\bra{\psi}\mathbb{I}\otimes\mathbb{I}-\mathbb{P}_{i_{0}i_{0}'}^{+}\otimes\mathbb{I}\otimes\ldots\otimes\mathbb{I}\ket{\psi}\ket{\psi}.
\]
Direct computation based on (\ref{eq:shmidt form}) shows that $\bra{\psi}\bra{\psi}\mathbb{P}_{i_{0}i_{0}'}^{+}\otimes\mathbb{I}\otimes\ldots\otimes\mathbb{I}\ket{\psi}\ket{\psi}<1$
which implies $\bra{\psi}\bra{\psi}\mathbb{I}\otimes\mathbb{I}-\mathbb{P}_{11'}^{+}\circ\mathbb{P}_{22'}^{+}\circ\ldots\circ\mathbb{P}_{LL'}^{+}\ket{\psi}\ket{\psi}>0$
. This concludes the proof of (\ref{eq:dist inf dim}).

\subsection{Bosons\label{sub:Bosons-inf}}

The criterion analogous to (\ref{eq:dist inf dim}) holds also for
the arbitrary finite number of bosonic particles with infinite dimensional
single particle Hilbert space. We have $\mathcal{H}_{b}=\mathrm{Sym}^{L}\left(\mathcal{H}\right)$,
where $\mathcal{H}$ is infinite dimensional. In analogy with the
finite dimensional case (\ref{bos}) we distinguish bosonic coherent
states:

\begin{equation}
\mathcal{O}_{b}=\left\{ \left[\ket{\phi}\otimes\ket{\phi}\otimes\ldots\otimes\ket{\phi}\right]|\,\ket{\phi}\in\mathcal{H}\right\} \,.
\end{equation}
We notice that coherent bosonic states are precisely completely symmetric
separable states of the system of identical distinguishable particles
with single particle Hilbert spaces $\mathcal{H}$. Thus, we can apply
criterion (\ref{eq:dist inf dim}) restricted to $\mathrm{Sym}^{L}\left(\mathcal{H}\right)$
to distinguish coherent bosonic states. More precisely, for $\ket{\psi}\in\mathrm{Sym}^{L}\left(\mathcal{H}\right)$
we have

\[
\left[\ket{\psi}\right]\in\mathcal{O}_{b}\Longleftrightarrow\bra{\psi}\bra{\psi}\mathbb{I}\otimes\mathbb{I}-\mathbb{P}_{11'}^{+}\circ\mathbb{P}_{22'}^{+}\circ\ldots\circ\mathbb{P}_{LL'}^{+}\ket{\psi}\ket{\psi}=0,
\]
where operator $\mathbb{P}_{11'}^{+}\circ\mathbb{P}_{22'}^{+}\circ\ldots\circ\mathbb{P}_{LL'}^{+}$
is assumed to act on the Hilbert space $\mathrm{Sym}^{L}\left(\mathcal{H}\right)\otimes\mathrm{Sym}^{L}\left(\mathcal{H}\right)\subset\mathcal{H}_{d}\otimes\mathcal{H}_{d}$,
with $\mathcal{H}_{d}$ defined in \ref{sub:Distinguishable-particles-inf}
and each single particle Hilbert space $\mathcal{H}_{i}$ equal to
$\mathcal{H}$.

\subsection{Fermions\label{sub:Fermions-inf}}

The case of fermionic particles turns out to be the most demanding,
albeit also the most interesting. System of $L$ fermionic particles
is described by $\mathcal{H}_{f}=\bigwedge^{L}\left(\mathcal{H}\right)$,
where the single particle Hilbert space $\mathcal{H}$ is infinite
dimensional. We define {}``non-entangled'' or coherent fermionic
states analogously to the finite dimensional case,

\begin{equation}
\mathcal{O}_{f}=\left\{ \left[\ket{\phi_{1}}\wedge\ket{\phi_{2}}\wedge\ldots\wedge\ket{\phi_{L}}\right]|\,\ket{\phi_{i}}\in\mathcal{H},\,\bk{\phi_{i}}{\phi_{j}}=\delta_{ij}\right\} \,.
\end{equation}
In what follows we prove that the criterion based on the generalized
concurrence (\ref{eq:conc ferm}) holds also in the infinite dimensional
situation. More precisely we show that

\begin{equation}
\left[\ket{\psi}\right]\in\mathcal{O}_{f}\Longleftrightarrow\bra{\psi}\bra{\psi}\mathbb{I}\otimes\mathbb{I}-\alpha\mathbb{P}_{11'}^{+}\circ\mathbb{P}_{22'}^{+}\circ\ldots\circ\mathbb{P}_{LL'}^{+}\ket{\psi}\ket{\psi}=0,\label{eq:ferm-crit- infinite}
\end{equation}
where, as before, $\alpha=\mbox{\ensuremath{\frac{2^{L}}{L+1}}}$.
It is assumed that $\mathbb{P}_{11'}^{+}\circ\mathbb{P}_{22'}^{+}\circ\ldots\circ\mathbb{P}_{LL'}^{+}$
acts on $\bigwedge^{L}\left(\mathcal{H}\right)\otimes\bigwedge^{L}\left(\mathcal{H}\right)\subset\mathcal{H}_{d}\otimes\mathcal{H}_{d}$,
with $\mathcal{H}_{d}$ defined in \ref{sub:Distinguishable-particles-inf}
and each single particle Hilbert space $\mathcal{H}_{i}$ equal to
$\mathcal{H}$. Let us denote $\mathbb{P}_{f}=\alpha\mathbb{P}_{11'}^{+}\circ\mathbb{P}_{22'}^{+}\circ\ldots\circ\mathbb{P}_{LL'}^{+}$.
In order to prove (\ref{eq:ferm-crit- infinite}) we consider the
equivalent problem,

\begin{equation}
\left[\ket{\psi}\right]\in\mathcal{O}_{f}\Longleftrightarrow\bra{\psi}\bra{\psi}\mathbb{P}_{f}\ket{\psi}\ket{\psi}=1\,,\label{eq:ferm-inf-form}
\end{equation}
for a normalized $\ket{\psi}\in\bigwedge^{L}\left(\mathcal{H}\right)$.
Note that if the rank of $\ket{\psi}$ (i.e. the minimal number of
elements of the form $\ket{\phi_{1}}\wedge\ket{\phi_{2}}\wedge\ldots\wedge\ket{\phi_{L}}$
needed to express $\ket{\psi}$) is finite, we have $\ket{\psi}\in\bigwedge^{L}\left(\mathcal{H}_{0}\right)$,
where $\mathcal{H}_{0}$ is some finite dimensional subspace of $\mathcal{H}$.
Therefore, in this case (\ref{eq:ferm-inf-form}) is proven as we
can apply results from Section \ref{sec:Nonlinear-entanglement-witness}
and Part \ref{sub:Fermions}. If rank of $\ket{\psi}$ is infinite
and $\bra{\psi}\bra{\psi}\mathbb{P}_{f}\ket{\psi}\ket{\psi}<1$ there
is nothing to prove. The only case left is when $\ket{\psi}$ has
infinite rank and $\bra{\psi}\bra{\psi}\mathbb{P}_{f}\ket{\psi}\ket{\psi}=1$.
In Appendix B we show that such a situation is impossible.

\section{Entanglement detection for mixed bosonic and fermionic states\label{sec:Entanglement-detection-for}}

From now on we will assume that dimensions of Hilbert spaces we consider
are finite. We will use notation introduced in Sections \ref{sec:Nonlinear-entanglement-witness}
and \ref{sec:Explicit-expressions-for} Recall that a mixed state
of a quantum system described by a Hilbert space $\mathcal{H}$ is
any operator $\rho$ on $\mathcal{H}$ satisfying conditions $\rho\geq0$
and $\mathrm{tr}\rho=1$. 

We treat separable mixed states for distinguishable particles, bosons
and fermions in the unified fashion. Let $K$ be a semisimple compact
Lie group irreducibly represented in the Hilbert space $\mathcal{H}$.
We say that a mixed state $\rho\in End\left(\mathcal{H}\right)$ (by
$End\left(\mathcal{H}\right)$ we denote the set of all operators
on $\mathcal{H}$) is a {}``generalized separable'' or {}``quasi-classical''
\cite{classical quantum states,On detection} if and only if $\rho$
can be written as a convex combination of projectors onto coherent
states of the action of $K$, i.e.

\begin{equation}
\rho=\sum_{i}p_{i}\kb{\psi_{i}}{\psi_{i}},\label{eq:coh mixed}
\end{equation}
where $p_{i}>0$, $\sum_{i}p_{i}=1$, and $\ket{\psi_{i}}$ are normalized
representatives of separable states $\left[\ket{\psi_{i}}\right]\in\mathcal{O}_{0}$
(see (\ref{eq:coherent})). By the appropriate choice of the group
and its representation, as discussed in Section \ref{sec:Nonlinear-entanglement-witness},
one recovers usual definition of mixed separable, as well as mixed
coherent bosonic and mixed fermionic states \cite{Kotowscy}. The
problem of deciding whether a given mixed state $\rho$ is coherent
is in general very difficult as in the decomposition (\ref{eq:coh mixed})
vectors $\ket{\psi_{i}}$ need not to be orthogonal. One way to solve
this problem, at least in principle, is to compute the convex roof
extension \cite{ulhman} of the generalized concurrence $C\left(\ket{\psi}\right)$
(\ref{eq:concurrence}). It is defined by

\begin{equation}
C\left(\rho\right)=\inf_{\sum_{i}p_{i}\kb{\phi_{i}}{\phi_{i}}=\rho}\left(\sum_{i}p_{i}C\left(\ket{\phi_{i}}\right)\right),\label{eq:convex roof}
\end{equation}
where the infimum is taken over all possible presentations of $\rho$
as a convex sum of one dimensional projectors. If dimension of $\mathcal{H}$
is finite the above expression is well defined because $C\left(\ket{\phi_{i}}\right)$
is a continuous function \cite{ulhman}. Moreover, $C\left(\rho\right)=0$,
if and only if $\rho$ is a coherent state. In what follows we denote
by $C_{d}\left(\rho\right)$, $C_{b}\left(\rho\right)$ and $C_{f}\left(\rho\right)$
convex roof extensions of generalized concurrences for distinguishable
particles, bosons and fermions. The explicit form of the function
$C\left(\rho\right)$ is in general not known (see \cite{On detection}
for the general discussion of situations in which $C\left(\rho\right)$
can be explicitly computed). For this reason it is desirable to have
non-trivial lower bounds for $C\left(\rho\right)$ that are easy to
compute. Such lower bounds give necessary conditions for a given mixed
state to be entangled. The problem of finding lower bounds for the
concurrence has been intensively studied for distinguishable particles
\cite{Observable bound,scalable,Agu=00015Bciak}. Due to the possible
experimental application, lower bounds for $C\left(\rho\right)$ that
can be expressed as the expectation value of some observable on multiple
copies of the state considered, are of greatest interest. An important
lower bound for the system of $L$ distinguishable particles (\ref{eq:dist})
is the so-called Mintert-Buchleitner bound \cite{scalable},

\begin{equation}
C_{d}\left(\rho\right)^{2}\geq\mathrm{tr}\left(\rho\otimes\rho V\right),\label{eq:eperimental bound dist}
\end{equation}
where operator $V:\mathcal{H}_{d}\otimes\mathcal{H}_{d}\rightarrow\mathcal{H}_{d}\otimes\mathcal{H}_{d}$
is given by the formula 
\[
V=\mathbb{I}\otimes\mathbb{I}-\mathbb{P}_{11'}^{+}\circ\mathbb{P}_{22'}^{+}\circ\ldots\circ\mathbb{P}_{LL'}^{+}-2\left(1-2^{-L}\right)\mathbb{P}^{-}.
\]
Operator $\mathbb{P}^{-}$ in the above expression denotes the projection
onto $\bigwedge^{2}\left(\mathcal{H}_{d}\right)$, the asymmetric
subspace of $\mathcal{H}_{d}\otimes\mathcal{H}_{d}$ ~. A necessary
condition for a sate $\rho$ to be entangled is thus $\mathrm{tr}\left(\rho\otimes\rho V\right)>0$.
In what follows we present the systematic way to construct lower bounds
for the generalized concurrences for bosons and fermions starting
from any lower bound for the concurrence for distinguishable particles.
Proofs of our results rely on the similarity of $C_{d}\left(\ket{\psi}\right)$
with $C_{b}\left(\ket{\psi}\right)$ and $C_{f}\left(\ket{\psi}\right)$
(compare (\ref{eq:conc disting}) with (\ref{eq:conc bos}) and (\ref{eq:conc ferm}))
and the fact that we can embed $\mathcal{H}_{b}$ and $\mathcal{H}_{f}$
in $\mathcal{H}_{d}$ .

\subsection{Bosons}

The generalized concurrence for mixed states of $L$ bosons reads,

\begin{equation}
C_{b}\left(\rho\right)=\inf_{\sum_{i}p_{i}\kb{\phi_{i}^{s}}{\phi_{i}^{s}}=\rho}\left(\sum_{i}p_{i}C_{b}\left(\ket{\phi_{i}^{s}}\right)\right)=\inf_{\sum_{i}p_{i}\kb{\phi_{i}^{s}}{\phi_{i}^{s}}=\rho}\left(\sum_{i}p_{i}C_{d}\left(\ket{\phi_{i}^{s}}\right)\right),\label{eq:mixed con bos}
\end{equation}
where the infimum is over all possible presentations of $\rho$ as
a convex sum of one dimensional projectors onto normalized $\ket{\phi_{i}^{s}}\in\mathrm{Sym}^{L}\left(\mathcal{H}\right)$.
In the second equality we used the fact $C_{b}\left(\ket{\phi_{i}^{s}}\right)=C_{d}\left(\ket{\phi_{i}^{s}}\right)$
for $\ket{\phi_{i}^{s}}\in\mathrm{Sym}^{L}\left(\mathcal{H}\right)$.
Due to (\ref{eq:mixed con bos}), we have: $C_{b}\left(\rho\right)\geq C_{d}\left(\rho\right)$
for $\rho\in\mathrm{Sym}^{L}\left(\mathrm{\mathcal{H}}\right)$. Indeed, 

\begin{equation}
C_{b}\left(\rho\right)=\inf_{\sum_{i}p_{i}\kb{\phi_{i}^{s}}{\phi_{i}^{s}}=\rho}\left(\sum_{i}p_{i}C_{d}\left(\ket{\phi_{i}^{s}}\right)\right)\geq\inf_{\sum_{i}p_{i}\kb{\phi_{i}}{\phi_{i}}=\rho}\left(\sum_{i}p_{i}C_{d}\left(\ket{\phi_{i}}\right)\right)=C_{d}\left(\rho\right),
\end{equation}
where the second infimum is over all presentations of the state $\rho$
as a convex sum of one dimensional projectors onto normalized $\ket{\phi_{i}}\in\mathcal{H}_{d}=\otimes^{L}\left(\mathcal{H}\right)$.
As a result, any lower bound for the concurrence for distinguishable
particles, i.e. a function on mixed states satisfying $C_{d}\left(\rho\right)\geq f\left(\rho\right)$,
is a lower bound for concurrence for bosons, $C_{b}\left(\rho\right)\geq f\left(\rho\right)$.
In particular we have,

\begin{equation}
C_{b}\left(\rho\right)^{2}\geq\mathrm{tr}\left(\rho\otimes\rho\tilde{V}\right),\label{eq:mintert bos}
\end{equation}
where $\tilde{V}=\left.V\right|_{\mathrm{\bigwedge}^{L}\left(\mathcal{H}\right)\otimes\mathrm{\bigwedge}^{L}\left(\mathcal{H}\right)}$
(\ref{eq:eperimental bound dist}). In general one can try to exploit
the additional symmetries of the bosonic Hilbert space $\mathrm{Sym}^{L}\left(\mathcal{H}\right)$
to get improved lower bounds for $C_{b}\left(\rho\right)$ but we
do not address this problem here.

\subsection{Fermions}

The generalized concurrence for mixed states of $L$ fermions is given
by

\begin{equation}
C_{f}\left(\rho\right)=C_{\alpha}\left(\rho\right)=\inf_{\sum_{i}p_{i}\kb{\phi_{i}^{a}}{\phi_{i}^{a}}=\rho}\left(\sum_{i}p_{i}C_{f}\left(\ket{\phi_{i}^{a}}\right)\right),\label{eq:mixed conc ferm}
\end{equation}
where the infimum is over all possible presentations of the state
$\rho$ as a convex sum of one dimensional projectors onto normalized
$\ket{\phi_{i}^{a}}\in\mathrm{\bigwedge}^{L}\left(\mathcal{H}\right)$.
We wrote $C_{\alpha}\left(\rho\right)$ to indicate the dependence
on the number $\alpha\geq1$ (itself depending upon the number of
particles $L$) that appears in (\ref{eq:conc ferm}). We have the
following inequality

\begin{equation}
C_{\alpha}\left(\rho\right)\geq\sqrt{\alpha}C_{d}\left(\rho\right)-\sqrt{\alpha-1}\,.\label{eq:conc bound fermions}
\end{equation}
The proof of the above relies on the relation between $C_{d}\left(\ket{\psi}\right)$
with $C_{f}\left(\ket{\psi}\right)$ (see (\ref{eq:conc disting})
and (\ref{eq:conc ferm})). For a normalized vector $\ket{\psi^{a}}\in\bigwedge^{L}\left(\mathcal{H}\right)$
we have 

\begin{equation}
C_{d}\left(\ket{\psi^{a}}\right)\geq\frac{1}{\sqrt{\alpha}}C_{\alpha}\left(\ket{\psi^{a}}\right)+\sqrt{1-\frac{1}{\alpha}}\,.\label{eq:pure inequality}
\end{equation}
Indeed,

\begin{align}
C_{d}\left(\ket{\psi^{a}}\right)= & \sqrt{\bra{\psi^{a}}\bra{\psi^{a}}\mathbb{I}\otimes\mathbb{I}-\frac{1}{\alpha}\mathbb{P}_{f}\ket{\psi^{a}}\ket{\psi^{a}}}=\sqrt{\frac{1}{\alpha}\bra{\psi^{a}}\bra{\psi^{a}}\mathbb{I}\otimes\mathbb{I}-\mathbb{P}_{f}\ket{\psi^{a}}\ket{\psi^{a}}+\left(1-\frac{1}{\alpha}\right)}\nonumber \\
 & \le\frac{1}{\sqrt{\alpha}}\sqrt{\bra{\psi^{a}}\bra{\psi^{a}}\mathbb{I}\otimes\mathbb{I}-\mathbb{P}_{f}\ket{\psi^{a}}\ket{\psi^{a}}}+\sqrt{1-\frac{1}{\alpha}}=\frac{1}{\sqrt{\alpha}}C_{\alpha}\left(\ket{\psi^{a}}\right)+\sqrt{1-\frac{1}{\alpha}}\,,
\end{align}
where we have used (\ref{eq:conc disting}) and (\ref{eq:conc ferm})
and inequality $\sqrt{a+b}\leq\sqrt{a}+\sqrt{b}$ for $a,\, b\geq0$.
For a given $\rho$ we apply to  (\ref{eq:pure inequality}) the operation
of convex roof extension (\ref{eq:mixed conc ferm}). We get the inequality,

\begin{equation}
C_{\alpha}\left(\rho\right)+\sqrt{\alpha-1}\geq\sqrt{\alpha}\inf_{\sum_{i}p_{i}\kb{\phi_{i}^{a}}{\phi_{i}^{a}}=\rho}\left(\sum_{i}p_{i}C_{d}\left(\ket{\phi_{i}^{a}}\right)\right).
\end{equation}
We conclude the proof of (\ref{eq:conc bound fermions}) by noting
that 

\[
\inf_{\sum_{i}p_{i}\kb{\phi_{i}^{a}}{\phi_{i}^{a}}=\rho}\left(\sum_{i}p_{i}C_{d}\left(\ket{\phi_{i}^{a}}\right)\right)\geq\inf_{\sum_{i}p_{i}\kb{\phi_{i}}{\phi_{i}}=\rho}\left(\sum_{i}p_{i}C_{d}\left(\ket{\phi_{i}^{a}}\right)\right)=C_{d}\left(\rho\right)\,,
\]
where the second infimum is over all presentations of $\rho$ as a
convex sum of one dimensional projectors onto normalized $\ket{\phi_{i}}\in\mathcal{H}_{d}=\otimes^{L}\left(\mathcal{H}\right)$. 

Consider any lower bound $f\left(\rho\right)$ for the concurrence
for $L$ distinguishable particles. From inequality (\ref{eq:conc bound fermions})
it follows that we have

\begin{equation}
C_{\alpha}\left(\rho\right)\geq\sqrt{\alpha}f\left(\rho\right)-\sqrt{\alpha-1}\,,
\end{equation}
where $f\left(\rho\right)$ is restricted to operators on $\mathrm{\bigwedge}^{L}\left(\mathcal{H}\right)$.
Therefore, condition $\sqrt{\alpha}f\left(\rho\right)-\sqrt{\alpha-1}>0$
gives a necessary condition fro a mixed fermionic state to be entangled.
Assume now that for $L$ distinguishable particles we have inequality,

\begin{equation}
C_{d}\left(\rho\right)^{2}\geq\mathrm{tr}\left(\rho\otimes\rho V\right)\,,
\end{equation}
for some operator $V$ on $\mathcal{H}_{d}$. From (\ref{eq:conc bound fermions})
we have

\[
C_{\alpha}^{2}\left(\rho\right)+2\sqrt{\alpha-1}C_{\alpha}\left(\rho\right)+\alpha-1\geq\alpha C_{d}^{2}\left(\rho\right)\,.
\]
It follows that

\[
C_{\alpha}\left(\rho\right)\left(1+2\sqrt{\alpha-1}C_{\alpha}\left(\rho\right)\right)\geq\alpha C_{d}^{2}\left(\rho\right)-\left(\alpha-1\right)\,.
\]
Application of (\ref{eq:eperimental bound dist}) to the above formula
results in the inequality, 

\begin{equation}
C_{\alpha}\left(\rho\right)\left(1+2\sqrt{\alpha-1}C_{\alpha}\left(\rho\right)\right)\geq\mathrm{tr}\left(\rho\otimes\rho\tilde{V}\right),\label{eq:frm eperim bound}
\end{equation}
where $\tilde{V}=\alpha V-\left(\alpha-1\right)\mathbb{I}\otimes\mathbb{I}$
and acts on $\mathrm{\bigwedge}^{L}\left(\mathcal{H}\right)\otimes\mathrm{\bigwedge}^{L}\left(\mathcal{H}\right)$.
From (\ref{eq:frm eperim bound}) it follows that $\mathrm{tr}\left(\rho\otimes\rho\tilde{V}\right)>0$
is a necessary condition for a given fermionic state $\rho$ to be
{}``entangled'' as $C_{\alpha}\left(\rho\right)\left(1+2\sqrt{\alpha-1}C_{\alpha}\left(\rho\right)\right)>0$
if and only if $C_{\alpha}\left(\rho\right)>0$. Application of the
above reasoning to the Mintert-Buchleitner bound \ref{eq:frm eperim bound}
gives a particularly simple expression for $\tilde{V}$,

\begin{align}
\tilde{V} & =\alpha\left(\mathbb{I}\otimes\mathbb{I}-\mathbb{P}_{11'}^{+}\otimes\mathbb{P}_{11'}^{+}\circ\mathbb{P}_{22'}^{+}\circ\ldots\circ\mathbb{P}_{LL'}^{+}-2\left(1-2^{-L}\right)\mathbb{P}^{-}\right)-\left(\alpha-1\right)\mathbb{I}\otimes\mathbb{I}\,,\\
 & =\left(\mathbb{I}\otimes\mathbb{I}-\mathbb{P}_{f}\right)-2\cdot\alpha\left(1-2^{-L}\right)\mathbb{P}^{-}
\end{align}
where $\mathbb{P}^{-}$ is the projector onto $\bigwedge^{2}\left(\bigwedge^{L}\left(\mathcal{H}\right)\right)\subset\mathrm{\bigwedge}^{L}\left(\mathcal{H}\right)\otimes\mathrm{\bigwedge}^{L}\left(\mathcal{H}\right)$.

\section{Summary}

We presented a comprehensive discussion of generalized concurrence
which is an extension of the usual concurrence to systems consisting
of not only distinguishable but also non-distinguishable particles.
The generalized concurrence can be used to detect non-coherent or
entangled pure states of bosonic or fermionic systems. Using tools
of representation theory we gave a closed form expressions for concurrences
for systems consisting of arbitrary, albeit finite, number of distinguishable
particles (\ref{eq:conc disting}), bosons (\ref{eq:conc bos}) or
fermions (\ref{eq:conc ferm}). We proved that expressions defining
concurrences are valid also when single particle Hilbert spaces are
infinite dimensional (Section \ref{sec:Generalization-to-infinite}).
In the last part of the article we studied mixed states of bosons
and fermions with the help of convex roof extensions of appropriate
concurrences, $C_{b}\left(\rho\right)$ and $C_{f}\left(\rho\right)$.
We used the connection between concurrences for distinguishable and
non-distinguishable particles to obtain lower bounds for $C_{b}\left(\rho\right)$
and $C_{f}\left(\rho\right)$ from any lower bound for the concurrence
for distinguishable particles, $C_{d}\left(\rho\right)$. This approach
allowed us to obtain non-trivial lower bounds for $C_{b}^{2}\left(\rho\right)$
and $C_{f}^{2}\left(\rho\right)$ in the spirit of Mintert-Buchleitner
(see (\ref{eq:mintert bos}) and (\ref{eq:frm eperim bound})).

\section{Acknowledgments}

We would like to thank Florian Mintert for fruitful discussions. We
gratefully acknowledge the support of of the ERC grant QOLAPS.

\section*{Appendix A Finite dimensional case\label{sec:Apppendix 1}}

\subsection*{Representation theory of $SU(N)$}

Representation theory of semisimple Lie groups and algebras is a rich
and beautiful but we will not discuss it here. We refer interested
reader to the relevant literature of the subject \cite{Hall,Barut Raczka}.
For a more elaborate discussion of representation-theoretic methods
in the context of entanglement theory see \cite{Kotowscy,On detection}.
In this section we briefly describe basic facts from representation
theory of $SU(N)$. The group $SU(N)$ is an example of a semisimple
Lie group whose representation theory exhibits essentially all the
features the general theory.

Let $\mathfrak{su}\left(N\right)$ be a Lie algebra of $SU(N)$, i.e.
a real Lie algebra consisting of all skew-hermitian and traceless
$N\times N$ matrices. It is useful to study the complexifications
\cite{Hall} of the group $SU(N)$ and its Lie algebra. The complexified
group $SU(N)^{\mathbb{C}}=SL\left(N\right)$, consists of all complex
$N\times N$ matrices with determinant $1$. The complexified algebra
$\mathfrak{su}\left(N\right)^{\mathbb{C}}=\mathfrak{sl}\left(N\right)$
consists of all complex traceless $N\times N$ matrices with zero
trace. Irreducible representations of $SU(N)$, $\mathfrak{su}(N)$,
$SL(N)$ and $\mathfrak{sl}(N)$ are in one to one correspondence
- if $\mathcal{H}$ is an irreducible representation of any of four
structures specified, then it is necessary an irreducible representation
of the remaining three structures. Lie algebra $\mathfrak{sl}(N)$
turns is particularly useful in the description of irreducible representations
of $SU(N).$ We have the following decomposition of $\mathfrak{sl}(N)$:
\[
\mathfrak{sl}(N)=\mathfrak{n}_{-}\oplus\mathfrak{h}\oplus\mathfrak{n}_{+}\,,
\]
where $\mathfrak{h}$ consists of diagonal traceless matrices and
$\mathfrak{n}_{-}$ and $\mathfrak{n}_{+}$ are respectively strictly
lower and upper diagonal matrices. Let $\pi$ be the irreducible representation
of $\mathfrak{sl}(N)$ in the Hilbert space $\mathcal{H}$. A convenient
way of description of the representation $\pi$ uses the notion of
weights vectors, i.e., simultaneous eigenvectors of representatives
of all elements form the Cartan subalgebra $\mathfrak{h}$. It means
that $\ket{\psi_{\lambda}}\in\mathcal{H}$ is a weight vector if,

\begin{equation}
\pi(H)\ket{\psi_{\lambda}}=\lambda(H)\ket{\psi_{\lambda}}\,,
\end{equation}
 for $H\in\mathcal{\mathfrak{h}}$, where a form $\lambda\in\mathfrak{h}^{\ast}$
is called ta weight of $\pi$. We have the decomposition,

\begin{equation}
\mathcal{H}=\oplus_{\lambda}\mathcal{H}_{\lambda},
\end{equation}
where summation is over all weights of the considered representation.
The subspaces $\mathcal{H}_{\lambda}$ are spanned by vectors corresponding
to the corresponding weight $\lambda$. An irreducible representation
is uniquely characterized by its highest weight $\lambda_{0}$ determined
by the highest weight vector $\ket{\psi_{\lambda_{0}}}$, i.e. by
the (unique, up to the multiplicative constant) weight vector annihilated
by all representatives of $n_{+}$: 
\begin{equation}
\pi(H)\ket{\psi_{\lambda_{0}}}=\lambda_{0}(H)\,\text{for \ensuremath{H\in\mathfrak{h}}\,\ and }\pi(\mathfrak{n}_{+})\ket{\psi_{\lambda_{0}}}=0\,.\label{eq:highest weight}
\end{equation}
Given the highest weight vector $\ket{\psi_{\lambda_{0}}}$, we can
generate the whole $\mathcal{H}$ by the action of $\mathfrak{sl}(N)$
(or equivalently by the action of $\mathfrak{k}$, $K$ or $G$):
$\mathcal{H}=\mathrm{span}_{\mathbb{C}}\left\{ \pi(X)\ket{\psi_{\lambda_{0}}}|X\in\mathfrak{sl}(N)\right\} $.
We write $\mathcal{H}^{\lambda_{0}}$ instead of $\mathcal{H}$ when
we want to distinguish which irreducible representation of $\mathfrak{sl}(N)$
is considered.

\subsection*{Formulas for $\mathbb{P}^{2\lambda_{0}}$ }

In this part we prove formulas for $\mathbb{P}^{2\lambda_{0}}$ in
the case of distinguishable particles, bosons and fermions.

\subsubsection*{Distinguishable particles}

Let $\mathcal{H}^{\lambda_{0}}=\mathcal{H}_{d}=\bigotimes_{i=1}^{i=L}\mathcal{H}$,
$\mathcal{H}\approx\mathbb{C}^{N}$ and $K=\times_{i=1}^{i=L}SU(N)$
We show here that $\mathbb{P}_{d}:\bigotimes^{2L}\mathcal{H}\rightarrow\bigotimes^{2L}\mathcal{H}$
defined by 

\begin{equation}
\mathbb{P}_{f}=\mathbb{P}_{11'}^{+}\circ\mathbb{P}_{22'}^{+}\circ\ldots\circ\mathbb{P}_{LL'}^{+}\,,
\end{equation}
equals $\mathbb{P}^{2\lambda_{0}}$. The proof of this statement is
the following. First, notice that for separable $\ket{\psi}$ we have
$\mathbb{P}_{d}\ket{\psi}\otimes\ket{\psi}=\ket{\psi}\otimes\ket{\psi}$.
Secondly, notice that have the equivalence of representations of $K$,
\[
\mathbb{P}_{11'}^{+}\circ\mathbb{P}_{22'}^{+}\circ\ldots\circ\mathbb{P}_{LL'}^{+}\,\left(\mathrm{Sym}^{2}\left(\mathcal{H}_{d}\right)\right)\approx\mathrm{Sym}^{2}\left(\mathcal{H}_{1}\right)\otimes\mathrm{Sym}^{2}\left(\mathcal{H}_{2}\right)\otimes\ldots\otimes\mathrm{Sym}^{2}\left(\mathcal{H}_{L}\right)\,.
\]
Therefore, subspace $\mathbb{P}_{11'}^{+}\circ\mathbb{P}_{22'}^{+}\circ\ldots\circ\mathbb{P}_{LL'}^{+}\,\left(\mathrm{Sym}^{2}\left(\mathcal{H}_{d}\right)\right)$
is an irreducible representation of $K$. Talking into account criterion
(\ref{eq:class criterion}) and the fact that separable states are
exactly coherent states of $K$ finishes the proof.

\subsubsection*{Bosons }

Let$\mathcal{H}^{\lambda_{0}}=\mathcal{H}_{b}=\mathrm{Sym}^{L}\left(\mathcal{H}\right),\mathcal{H}\approx\mathbb{C}^{N}$
and $K=SU(N)$. We show that the operator $\mathbb{P}_{b}:\bigotimes^{2L}\mathcal{H}\rightarrow\bigotimes^{2L}\mathcal{H}$
given by

\begin{equation}
\mathbb{P}_{b}=\left(\mathbb{P}_{11'}^{+}\circ\mathbb{P}_{22'}^{+}\circ\ldots\circ\mathbb{P}_{LL'}^{+}\right)\left(\mathbb{P}_{\left\{ 1,\ldots,L\right\} }^{\mathrm{sym}}\circ\mathbb{P}_{\left\{ 1',\ldots,L'\right\} }^{\mathrm{sym}}\right)\,,\label{eq:bosproof-1}
\end{equation}
equals $\mathbb{P}^{2\lambda_{0}}.$ The proof is the following. Notice
that 
\begin{equation}
\mathbb{P}_{b}\left(\mathrm{Sym}^{L}\left(\mathcal{H}\right)\vee\mathrm{Sym}^{L}\left(\mathcal{H}\right)\right)\subset\mathrm{Sym}^{L}\left(\mathcal{H}\right)\vee\mathrm{Sym}^{L}\left(\mathcal{H}\right)\,.
\end{equation}
Moreover, $\mathbb{P}_{b}$ is a projector onto $\mathrm{Sym}^{2L}\left(\mathcal{H}\right)$,
a completely symmetric subspace of $\mathcal{H}_{d}\otimes\mathcal{H}_{d}$.
Subspace $\mathrm{Sym}^{2L}\left(\mathcal{H}\right)$ is an irreducible
representation of $K$. For a coherent bosonic state $\ket{\psi}\in\mathcal{H}_{b}$
, we have $\mathbb{P}_{b}\ket{\psi}\otimes\mathbb{\ket{\psi}}=\ket{\psi}\otimes\mathbb{\ket{\psi}}$.
As a result, by criterion (\ref{eq:class criterion}), $\mathbb{P}_{b}=\mathbb{P}^{2\lambda_{0}}$.

\subsubsection*{Fermions}

Let $\mathcal{H}^{\lambda_{0}}=\bigwedge^{L}\left(\mathcal{H}\right)$,
$\mathcal{H}\approx\mathbb{C}^{N}$, $K=SU(N)$ and $\alpha=\frac{2^{L}}{L+1}$
. We prove that $\mathbb{P}_{f}:\bigotimes^{2L}\mathcal{H}\rightarrow\bigotimes^{2L}\mathcal{H}$
defined by 

\begin{equation}
\mathbb{P}_{f}=\mbox{\ensuremath{\alpha}}\left(\mathbb{P}_{11'}^{+}\circ\mathbb{P}_{22'}^{+}\circ\ldots\circ\mathbb{P}_{LL'}^{+}\right)\left(\mathbb{P}_{\left\{ 1,\ldots,L\right\} }^{\mathrm{asym}}\circ\mathbb{P}_{\left\{ 1',\ldots,L'\right\} }^{\mathrm{asym}}\right)\label{eq:ferm proof}
\end{equation}
is precisely $\mathbb{P}^{2\lambda_{0}}$, the projector onto $\mathcal{H}^{2\lambda_{0}}\subset\bigwedge^{L}\left(\mathcal{H}\right)\otimes\bigwedge^{L}\left(\mathcal{H}\right)\subset\bigotimes^{2L}\mathcal{H}$
(consult Section \ref{sub:Fermions}). The full proof relies on the
representation theory of $SU(N)$. Main technical tools involved are
Young diagrams, Schur-Weyl duality and the theory of plethysms \cite{Barut Raczka,Cvitanovic,Young tableaux}.
In order to simplify the reasoning we base our argumentation on two
simple facts:
\begin{enumerate}
\item Operator $\mathbb{P}_{f}$ is the projector onto some irreducible
representation of $SU(N)$ in $\bigotimes^{2L}\mathcal{H}$.
\item $\mathbb{P}_{f}\left(\ket{\psi_{\lambda_{0}}}\otimes\ket{\psi_{\lambda_{0}}}\right)=\ket{\psi_{\lambda_{0}}}\otimes\ket{\psi_{\lambda_{0}}}$,
where $\ket{\psi_{\lambda_{0}}}=\ket{\psi_{1}}\wedge\ket{\psi_{2}}\wedge\ldots\wedge\ket{\psi_{L}}$
is the highest weight vector of the representation $\mathcal{H}^{\lambda_{0}}$.
\end{enumerate}
Proof of the Fact 1 can be found in \cite{Cvitanovic}. Before we
prove Fact 2 let us assume for the moment that above two facts are
true. Because $\mathbb{P}_{f}$ preserves $\ket{\psi_{\lambda_{0}}}\otimes\ket{\psi_{\lambda_{0}}}$and
from the vector $\ket{\psi_{\lambda_{0}}}\otimes\ket{\psi_{\lambda_{0}}}$
it is possible to generate (via the action of $SU(N)$) the whole
$\mathcal{H}^{2\lambda_{0}}\subset\bigwedge^{L}\left(\mathcal{H}\right)\otimes\bigwedge^{L}\left(\mathcal{H}\right)\subset\bigotimes^{2L}\mathcal{H}$,
one concludes that $\mathbb{P}_{f}=\mathbb{P}^{2\lambda_{0}}$. Let
us turn to the proof of the second fact. Let us fix the basis $\left\{ \ket{\psi_{i}}\right\} _{i=1}^{i=N}$
of $\mathcal{H}$ and let $\ket{\psi_{1}}\wedge\ket{\psi_{2}}\wedge\ldots\wedge\ket{\psi_{L}}=\ket{\psi_{\lambda_{0}}}$
be the (unnormalized) highest weight vector of the representation
$\bigwedge^{L}\left(\mathcal{H}\right)$. From the definition of the
wedge product we have 

\[
\mathbb{P}_{f}\left(\ket{\psi_{\lambda_{0}}}\otimes\ket{\psi_{\lambda_{0}}}\right)=\mathbb{P}_{f}\left(\ket{\psi_{1}}\wedge\ket{\psi_{2}}\wedge\ldots\wedge\ket{\psi_{L}}\otimes\ket{\psi_{1}}\wedge\ket{\psi_{2}}\wedge\ldots\wedge\ket{\psi_{L}}\right)\,,
\]

\[
=\mathbb{P}_{f}\left(\sum_{\sigma\in S_{L}}\sum_{\tau\in S_{L}}\mathrm{sgn\left(\sigma\right)}\mathrm{sgn}\left(\tau\right)\ket{\psi_{\sigma(1)}}\otimes\ket{\psi_{\sigma(2)}}\otimes\ldots\otimes\ket{\psi_{\sigma(L)}}\otimes\ket{\psi_{\tau(1)}}\otimes\ket{\psi_{\tau(2)}}\otimes\ldots\otimes\ket{\psi_{\tau(L)}}\right)\,
\]
\begin{equation}
=\frac{1}{L+1}\sum_{\sigma\in S_{L}}\sum_{\tau\in S_{L}}\mathrm{sgn\left(\sigma\tau\right)\left(\ket{\psi_{\sigma(1)}}\otimes\ket{\psi_{\tau(1)}}+\ket{\psi_{\tau(1)}}\otimes\ket{\psi_{\sigma(1)}}\right)}\otimes\ldots\otimes\left(\ket{\psi_{\sigma(L)}}\otimes\ket{\psi_{\tau(L)}}+\ket{\psi_{\tau(L)}}\otimes\ket{\psi_{\sigma(L)}}\right)\,.\label{eq:koszmar1}
\end{equation}
In the above expressions, $S_{L}$ denotes permutation group of $L$
elements and $\mathrm{sgn}\left(\cdot\right)$ denotes the sign of
a permutation. In order to simply the notation, we swapped order of
terms in the full tensor product $\bigotimes^{2L}\mathcal{H}$ i.e.
we used the isomorphism:

\[
\bigotimes^{2L}\mathcal{H}=\left(\bigotimes_{i=1}^{i=L}\mathcal{H}_{i}\right)\otimes\left(\bigotimes_{i=1'}^{i=L'}\mathcal{H}_{i}\right)\approx\left(\mathcal{H}_{1}\otimes\mathcal{H}_{1'}\right)\otimes\left(\mathcal{H}_{2}\otimes\mathcal{H}_{2'}\right)\otimes\ldots\otimes\left(\mathcal{H}_{L}\otimes\mathcal{H}_{L'}\right)\,,
\]
for $\mathcal{H}_{i}\approx\mathcal{H}.$ Let us introduce the notation

\[
\ket{\Phi_{k,\sigma,\theta}}=\left(\ket{\psi_{\tau(1)}}\otimes\ket{\psi_{\sigma(1)}}\right)\otimes\ldots\otimes\left(\ket{\psi_{\tau(k)}}\otimes\ket{\psi_{\sigma(k)}}\right)\otimes\left(\ket{\psi_{\sigma(k+1)}}\otimes\ket{\psi_{\tau(k+1)}}\right)\otimes\ldots\otimes\left(\ket{\psi_{\sigma(L)}}\otimes\ket{\psi_{\tau(L)}}\right)+\,
\]

\[
+\left(\ket{\psi_{\sigma(1)}}\otimes\ket{\psi_{\tau(1)}}\right)\otimes\left(\ket{\psi_{\tau(2)}}\otimes\ket{\psi_{\sigma(2)}}\right)\otimes\ldots\otimes\left(\ket{\psi_{\tau(k+1)}}\otimes\ket{\psi_{\sigma(k+1)}}\right)\otimes\left(\ket{\psi_{\sigma(k+2)}}\otimes\ket{\psi_{\tau(k+2)}}\right)\otimes\ldots+\ldots
\]
where $\ldots$ denotes the summation over remaining $\binom{L}{k}-2$
terms one obtains by the different choice of $k$ element combinations
from $\left\{ 1,\ldots,L\right\} $. Reordering of terms in (\ref{eq:koszmar1})
gives

\begin{equation}
\frac{1}{L+1}\sum_{k=0}^{k=L}\left(\sum_{\sigma\in S_{L}}\sum_{\tau\in S_{L}}\mathrm{sgn}\left(\sigma\tau\right)\ket{\Phi_{k,\sigma,\theta}}\right)\,.\label{eq:koszmar2}
\end{equation}
Operator $\mathbb{P}_{f}$ preserves $\bigwedge^{L}\left(\mathcal{H}\right)\otimes\bigwedge^{L}\left(\mathcal{H}\right)$
and therefore 
\[
\mathbb{P}_{f}\left(\ket{\psi_{\lambda_{0}}}\otimes\ket{\psi_{\lambda_{0}}}\right)=\left(\mathbb{P}_{\left\{ 1,\ldots,L\right\} }^{\mathrm{asym}}\circ\mathbb{P}_{\left\{ 1',\ldots,L'\right\} }^{\mathrm{asym}}\right)\circ\mathbb{P}_{f}\left(\ket{\psi_{\lambda_{0}}}\otimes\ket{\psi_{\lambda_{0}}}\right)\,.
\]
As a result from (\ref{eq:koszmar2}) we have

\begin{equation}
\frac{1}{L+1}\sum_{k=0}^{k=L}\left(\sum_{\sigma\in S_{L}}\sum_{\tau\in S_{L}}\mathrm{sgn}\left(\sigma\tau\right)\left(\mathbb{P}_{\left\{ 1,\ldots,L\right\} }^{\mathrm{asym}}\circ\mathbb{P}_{\left\{ 1',\ldots,L'\right\} }^{\mathrm{asym}}\right)\ket{\Phi_{k,\sigma,\theta}}\right)\,.\label{eq:koszmar3}
\end{equation}
We claim that for each $k=0,\ldots L$ we have
\begin{equation}
\sum_{\sigma\in S_{L}}\sum_{\tau\in S_{L}}\mathrm{sgn}\left(\sigma\tau\right)\left(\mathbb{P}_{\left\{ 1,\ldots,L\right\} }^{\mathrm{asym}}\circ\mathbb{P}_{\left\{ 1',\ldots,L'\right\} }^{\mathrm{asym}}\right)\left(\ket{\Phi_{k,\sigma,\theta}}\right)=\ket{\psi_{\lambda_{0}}}\otimes\ket{\psi_{\lambda_{0}}}\,.\label{eq:koszmar main}
\end{equation}
Indeed, application of $\mathbb{P}_{\left\{ 1,\ldots,L\right\} }^{\mathrm{asym}}\circ\mathbb{P}_{\left\{ 1',\ldots,L'\right\} }^{\mathrm{asym}}$
gives

\begin{align}
\frac{1}{\left(L!\right)^{2}}\sum_{\sigma\in S_{L}}\sum_{\tau\in S_{L}}\mathrm{sgn}\left(\sigma\tau\right)\left(\left(\ket{\psi_{\tau(1)}}\wedge\ket{\psi_{\tau(2)}}\wedge\ldots\wedge\ket{\psi_{\tau(k)}}\wedge\ket{\psi_{\sigma(k+1)}}\wedge\ldots\right)\right.\otimes\,\label{eq:koszmar4}\\
\ldots\left.\otimes\left(\ket{\psi_{\sigma(1)}}\wedge\ket{\psi_{\sigma(2)}}\wedge\ldots\wedge\ket{\psi_{\sigma(k)}}\wedge\ket{\psi_{\tau(k+1)}}\wedge\ldots\right)+\ldots\right)\,,\nonumber 
\end{align}
where $\ldots$ denotes the summation over remaining $\binom{L}{k}-1$
terms. Let $S_{L}\left(\sigma,\, k\right)$ denote the subgroup of
$S_{L}$ consisting of permutations that do not mix sets $\left\{ \sigma(1),\ldots,\sigma(k)\right\} $
and $\left\{ \sigma(k+1),\ldots,\sigma(L)\right\} $. We have $S_{L}\left(\sigma,\, k\right)\approx S_{k}\times S_{L-k}$.
As a result, for the fixed $\sigma\in S_{k}$ we have

\textbf{\footnotesize 
\[
\sum_{\tau\in S_{L}}\mathrm{sgn}\left(\sigma\tau\right)\left(\ket{\psi_{\tau(1)}}\wedge\ket{\psi_{\tau(2)}}\wedge\ldots\wedge\ket{\psi_{\tau(k)}}\wedge\ket{\psi_{\sigma(k+1)}}\wedge\ldots\right)\otimes\left(\ket{\psi_{\sigma(1)}}\wedge\ket{\psi_{\sigma(2)}}\wedge\ldots\wedge\ket{\psi_{\sigma(k)}}\wedge\ket{\psi_{\tau(k+1)}}\wedge\ldots\right)
\]
}{\footnotesize \par}

\textbf{\footnotesize 
\[
=\sum_{\tau\in S_{L}\left(\sigma,k\right)}\mathrm{sgn}\left(\sigma\tau\right)\mathrm{sgn\left(\tau\sigma^{-1}\right)}\left(\ket{\psi_{\sigma(1)}}\wedge\ket{\psi_{\sigma(2)}}\wedge\ldots\wedge\ket{\psi_{\sigma(L)}}\right)\otimes\left(\ket{\psi_{\sigma(1)}}\wedge\ket{\psi_{\sigma(2)}}\wedge\ldots\wedge\ket{\psi_{\sigma(L)}}\right)=\left(L-k\right)!\cdot k!\ket{\psi_{\lambda_{0}}}\otimes\ket{\psi_{\lambda_{0}}}\,.
\]
}Treating all other terms in the outer bracket of (\ref{eq:koszmar4})
in the similar fashion gives

\[
\frac{1}{\left(L!\right)^{2}}\left(\sum_{\sigma\in S_{L}}\binom{L}{k}\left(L-k\right)!\cdot k!\right)\ket{\psi_{\lambda_{0}}}\otimes\ket{\psi_{\lambda_{0}}}=\ket{\psi_{\lambda_{0}}}\otimes\ket{\psi_{\lambda_{0}}}\,,
\]
which proves (\ref{eq:koszmar main}). From (\ref{eq:koszmar main})
and (\ref{eq:koszmar2}) we conclude the proof of the second Fact
and therefore prove that $\mathbb{P}_{f}=\mathbb{P}^{2\lambda_{0}}$.

\section*{Appendix B \label{sec:Appendix-2}Infinite dimensional case}

\subsection*{Distinguishable particles}

We prove here that the state $\left[\ket{\psi}\right]\in\mathbb{P}\mathcal{H}_{d}$
is separable if and only if $\left[\ket{\psi}\right]\in\mathcal{O}_{sep}^{i}$
for $i=1,\ldots,L$ (for definition of $\mathcal{O}_{sep}^{i}$ see
(\ref{eq:sep part})). First note that a separable state $\left[\ket{\psi}\right]$
clearly belongs to $\mathcal{O}_{sep}^{i}$. On the other hand, if
$\left[\ket{\psi}\right]\in\mathcal{O}_{sep}^{i}$ , then $\ket{\psi}$
is an eigenvector (with eigenvalue $1$) of the operator 

\begin{equation}
\kb{\phi_{i}}{\phi_{i}}\otimes\mathbb{I}_{i},\label{eq:proj 1}
\end{equation}
where $\ket{\phi_{i}}\in\mathcal{H}_{i}$ and $\mathbb{I}_{i}$ is
the identity operator on $\left(\bigotimes_{j\neq i}\mathcal{H}_{j}\right)$.
Note that in order to do not complicate the notation in (\ref{eq:proj 1})
we do not respect the order of terms in the tensor product $\bigotimes_{i=1}^{i=L}\mathcal{H}_{i}$.
Note that we can repeat the above reasoning for all other $i=1,\ldots,L$.
As a result, we get that $\ket{\psi}$ is an eigenvector with the
eigenvalue $1$ of the operator 

\[
\mathbb{P}_{\ket{\psi}}=\kb{\phi_{1}}{\phi_{1}}\otimes\kb{\phi_{2}}{\phi_{2}}\otimes\ldots\otimes\kb{\phi_{L}}{\phi_{L}},
\]
where $\ket{\psi_{i}}\in\mathcal{H}_{i}$. Operator $\mathbb{P}_{\ket{\psi}}$
is a projector onto a separable state which concludes the proof that
$\left[\ket{\psi}\right]$ is separable.

\subsection*{Fermions}

We show here that a normalized fermionic state $\ket{\psi}\in\bigwedge^{L}\left(\mathcal{H}\right)$
having infinite rank cannot satisfy $\bra{\psi}\bra{\psi}\mathbb{P}_{f}\ket{\psi}\ket{\psi}=1$
(see (\ref{sub:Fermions-inf}) for the definition of $\mathbb{P}_{f}$).
In the course of argumentation we will need the fact that the set
of coherent fermionic states $\mathcal{O}_{f}$ is closed in $\mathbb{P}\mathcal{H}_{f}$.
We prove that $\mathcal{O}_{f}$ is closed directly from the definition.
Let $\left[\ket{\psi_{k}}\right]\in\mathcal{O}_{f}$ be a Cauchy sequence.
Let us fix $\epsilon>0$. Then for $n,m\,>n_{0}\left(\epsilon\right)$
we have 

\[
\mathrm{d}\left(\left[\ket{\psi_{n}}\right],\,\left[\ket{\psi_{m}}\right]\right)\leq\epsilon\,.
\]
Assuming that vectors $\ket{\psi_{k}}$ are normalized and making
use of (\ref{eq:metric structure}) we get 
\begin{equation}
\left|\bk{\psi_{n}}{\psi_{m}}\right|^{2}\geq1-\frac{\epsilon^{2}}{2}\,.\label{eq:estimate bef}
\end{equation}
Vectors $\ket{\psi_{n}}$ and $\ket{\psi_{m}}$ can be represented
(non uniquely) by Slater determinants 

\[
\ket{\psi_{n}}=\ket{\phi_{1}^{n}}\wedge\ket{\phi_{2}^{n}}\wedge\ldots\wedge\ket{\phi_{L}^{n}},\,\ket{\psi_{m}}=\ket{\phi_{1}^{m}}\wedge\ket{\phi_{2}^{m}}\wedge\ldots\wedge\ket{\phi_{L}^{m}}\,.
\]
Straightforward cancellations show that

\[
\bk{\psi_{n}}{\psi_{m}}=\mathrm{det}\left(M\right),\,
\]
where $M$ is $L\times L$ density matrix whose entries are given
by $M_{ij}=\bk{\phi_{i}^{n}}{\phi_{j}^{m}}$. By the appropriate choice
of the basis of $V_{n}$, subspace of $\bigwedge^{L}\left(\mathcal{H}\right)$
spanned by vectors $\ket{\phi_{1}^{n}},\ldots\ket{\phi_{L}^{m}}$matrix
$M$ can be made diagonal. That is we have

\[
\mathrm{det}\left(M\right)=\bk{\tilde{\phi}_{1}^{n}}{\phi_{1}^{m}}\cdot\ldots\cdot\bk{\tilde{\phi}_{L}^{n}}{\phi_{L}^{m}},
\]
where $\ket{\psi_{n}}=\ket{\tilde{\phi}_{1}^{n}}\wedge\ket{\tilde{\phi}_{2}^{n}}\wedge\ldots\wedge\ket{\tilde{\phi}_{L}^{n}}$.
Setting $\ket{\phi_{i}^{m}}=\ket{\tilde{\phi}_{i}^{m}}$ and talking
into account (\ref{eq:estimate bef}) we get

\[
\mathrm{max}_{i=1,\ldots,L}\left|\bk{\tilde{\phi}_{n}^{i}}{\tilde{\phi}_{m}^{i}}\right|^{2}\geq1-\frac{\epsilon^{2}}{2}\,,
\]
which means that for each $i=1,\ldots,L$ {}``single particle''
states $\left[\ket{\tilde{\phi}_{k}^{i}}\right]\in\mathbb{P}\mathcal{H}$
form a Cauchy sequence with respect to the metric (\ref{eq:metric structure})
. From the closedness of $\mathbb{P}\mathcal{H}$ we infer that for
each $i$ we have $\left[\ket{\tilde{\phi}_{i}^{k}}\right]\overset{k\rightarrow\infty}{\longrightarrow}\left[\ket{\tilde{\phi}_{i}^{\infty}}\right]\in\mathbb{P}\mathcal{H}$.
From that we conclude that 
\[
\left[\ket{\tilde{\phi}_{1}^{k}}\wedge\ket{\tilde{\phi}_{2}^{k}}\wedge\ldots\wedge\ket{\tilde{\phi}_{L}^{k}}\right]\overset{k\rightarrow\infty}{\longrightarrow}\left[\ket{\tilde{\phi}_{1}^{\infty}}\wedge\ket{\tilde{\phi}_{2}^{\infty}}\wedge\ldots\wedge\ket{\tilde{\phi}_{L}^{\infty}}\right]\,,
\]
which finishes the proof of closedness of $\mathcal{O}_{f}$. We can
now return to the original problem. We introduce the sequence of finite
dimensional subspaces 

\begin{equation}
\mathcal{H}_{1}\subset\mathcal{H}_{2}\subset\ldots\subset\mathcal{H}_{k}\subset\ldots\subset\mathcal{H}\,,\label{eq:sequance subspaces}
\end{equation}
such that $\bigcup_{l=1}^{l=\infty}\mathcal{H}_{i}=\mathcal{H}$.
To the above sequence we associate corresponding sequence of subspaces
of $\bigwedge^{L}\left(\mathcal{H}\right)$,

\begin{equation}
\bigwedge^{L}\left(\mathcal{H}_{1}\right)\subset\bigwedge^{L}\left(\mathcal{H}_{2}\right)\subset\ldots\subset\bigwedge^{L}\left(\mathcal{H}_{k}\right)\subset\ldots\subset\bigwedge^{L}\left(\mathcal{H}\right)\,.
\end{equation}
Obviously we have $\bigcup_{l=1}^{l=\infty}\bigwedge^{L}\left(\mathcal{H}_{l}\right)=\bigwedge^{L}\left(\mathcal{H}\right)$.
We fix the index $k$ and consider the following orthogonal splittings
of $\bigwedge^{L}\left(\mathcal{H}\right)$ and $\bigwedge^{L}\left(\mathcal{H}\right)\otimes\bigwedge^{L}\left(\mathcal{H}\right)$,

\begin{equation}
\bigwedge^{L}\left(\mathcal{H}\right)=\bigwedge^{L}\left(\mathcal{H}_{k}\right)\oplus\left[\bigwedge^{L}\left(\mathcal{H}_{k}\right)\right]^{\perp},\label{eq:single split}
\end{equation}

\begin{equation}
\bigwedge^{L}\left(\mathcal{H}\right)\otimes\bigwedge^{L}\left(\mathcal{H}\right)=\left[\bigwedge^{L}\left(\mathcal{H}_{k}\right)\otimes\bigwedge^{L}\left(\mathcal{H}_{k}\right)\right]\oplus\left[\bigwedge^{L}\left(\mathcal{H}_{k}\right)\otimes\bigwedge^{L}\left(\mathcal{H}_{k}\right)\right]^{\perp},\label{eq:double split}
\end{equation}
where orthogonal complements are taken with respect to the usual inner
products on $\bigwedge^{L}\left(\mathcal{H}\right)$ and $\bigwedge^{L}\left(\mathcal{H}\right)\otimes\bigwedge^{L}\left(\mathcal{H}\right)$
respectively. By $\mathbb{P}_{k}:\bigwedge^{L}\left(\mathcal{H}\right)\rightarrow\bigwedge^{L}\left(\mathcal{H}_{k}\right)$
we denote the orthogonal projector on $\bigwedge^{L}\left(\mathcal{H}_{k}\right)$.
We now prove that the infinite rank of $\ket{\psi}$ and $\bra{\psi}\bra{\psi}\mathbb{P}_{f}\ket{\psi}\ket{\psi}=1$
yield to the contradiction. Let us first note that for normalized
$\ket{\psi}$ condition $\bra{\psi}\bra{\psi}\mathbb{P}_{f}\ket{\psi}\ket{\psi}=1$
is equivalent to $\mathbb{P}_{f}\ket{\psi}\ket{\psi}=\ket{\psi}\ket{\psi}$.
Consider a decomposition

\begin{equation}
\ket{\psi}\ket{\psi}=\ket{\Psi_{k}}+\ket{\Psi_{k}^{\perp}},
\end{equation}
where $\ket{\Psi_{k}}\in\bigwedge^{L}\left(\mathcal{H}_{k}\right)\otimes\bigwedge^{L}\left(\mathcal{H}_{k}\right)$
and $\ket{\Psi_{k}^{\perp}}\in\left[\bigwedge^{L}\left(\mathcal{H}_{k}\right)\otimes\bigwedge^{L}\left(\mathcal{H}_{k}\right)\right]^{\perp}$.
We have $\mathbb{P}_{f}\ket{\psi}\ket{\psi}=\ket{\psi}\ket{\psi}$
and thus 

\begin{equation}
\mathbb{P}_{f}\ket{\Psi_{k}}+\mathbb{P}_{f}\ket{\Psi_{k}^{\perp}}=\ket{\Psi_{k}}+\ket{\Psi_{k}^{\perp}}\,.
\end{equation}
Because $\mathbb{P}_{f}$ preserves $\bigwedge^{L}\left(\mathcal{H}_{k}\right)\otimes\bigwedge^{L}\left(\mathcal{H}_{k}\right)$
we have $\bra{\Psi_{k}^{\perp}}\mathbb{P}_{f}\ket{\Psi_{k}}=0$ and
therefore $\mathbb{P}_{f}\ket{\Psi_{k}}=\ket{\Psi_{k}}$. Notice that
$\ket{\Psi_{k}}=\mathbb{P}_{k}\otimes\mathbb{P}_{k}\left(\ket{\psi}\ket{\psi}\right)$
and therefore $\ket{\Psi_{k}}$ is a product state. Because $\mathbb{P}_{f}\ket{\Psi_{k}}=\ket{\Psi_{k}}$
we see that $\mathbb{P}_{k}\ket{\psi}$ is actually an (non-normalized)
representative of some coherent fermionic state. We can repeat the
above construction for the arbitrary number $k$. We get

\begin{equation}
1=\lim_{k\rightarrow\infty}\bra{\psi}\mathbb{P}_{k}\ket{\psi},\,\label{eq:limit}
\end{equation}
where $\mathbb{P}_{k}\ket{\psi}$ is the (non-normalized) representative
of some coherent state. We therefore get $\left[\mathbb{P}_{k}\ket{\psi}\right]\overset{k\rightarrow\infty}{\rightarrow}\left[\ket{\psi}\right]$
(in a sense of (\ref{eq:metric structure})) . Since $\left[\mathbb{P}_{k}\ket{\psi}\right]\in\mathcal{O}_{f}$,
we get that $\left[\ket{\psi}\right]\in\mathcal{O}_{f}$ as the set
of coherent fermionic states $\mathcal{O}_{f}$ is closed in $\mathbb{P}\mathcal{H}_{f}$
. This is clearly in contradiction with the assumption that $\ket{\psi}$
has infinite rank.

\end{document}